\documentstyle[12pt,aasms4,epsfig]{article}
\def\simlt{\ \raise -2.truept\hbox{\rlap{\hbox{$\sim$}}\raise5.truept   
\hbox{$<$}\ }}                                                          %
\def\simgt{\ \raise -2.truept\hbox{\rlap{\hbox{$\sim$}}\raise5.truept   %
\hbox{$>$}\ }}                                                          %
\def\el{$e^{\pm}$ ~}
\def\be{\begin{equation}}
\def\ee{\end{equation}}
\def\newline{\hfil\break}
\begin{document}
\title{NEUTRALINOS AND THE ORIGIN OF RADIO HALOS IN 
CLUSTERS OF GALAXIES}
\author{S. COLAFRANCESCO}
\affil{Osservatorio Astronomico di Roma, Via Frascati, 33, I-00040\\
Monteporzio - ITALY\\
Email: cola@coma.mporzio.astro.it}

\and

\author{B. MELE}
\affil{INFN, Sezione di Roma, and University of Rome ``La Sapienza"\\
Rome - ITALY\\
Email: Barbara.Mele@roma1.infn.it}

\begin{abstract}
We assume that the supersymmetric lightest neutralino is 
a good candidate for the cold dark matter in the galaxy halo,
and  explore the possibility to produce extended diffuse  radio emission 
from high-energy electrons arising from the neutralino annihilation in 
galaxy clusters whose intracluster medium is filled
with a large-scale magnetic field.
We show that these electrons fit the population of seed relativistic electrons 
that is postulated  in many models for the origin of cluster 
radio halos.
If the magnetic field has a central value of $3 \div 30$ $\mu$G 
(depending on the Dark Matter profile) and is radially
decreasing from the cluster center, the population of 
seed relativistic electrons 
from neutralino annihilation can fit the radio halo spectra 
of two well studied cluster: Coma and 1E0657-56.
The shape and the frequency extension of the radio halo spectra 
are connected with the mass and physical composition of the neutralino.
A pure-gaugino neutralino with mass $M_{\chi} \geq 80$ GeV 
 can reasonably fit the radio halo spectra of 
both Coma and 1E0657-56.
This model provides a number of extra predictions that make it 
definitely testable.
On the one hand, it agrees quite well with the observations
that {\it (i)} the radio halo is centered on the cluster dynamical center, 
usually coincident with the center of its X-ray emission, 
{\it (ii)}   the radio halo surface brightness 
is similar to the X-ray one, and {\it (iii)} the monochromatic radio power at
$1.4$ GHz correlates strongly with the IC gas temperature.
On the other hand, the same model predicts that radio halos should be present 
in every cluster, which is not actually observed, 
although the predicted radio halo luminosities can change by factors up to 
$\sim 10^2 \div 10^6$,
depending on the amplitude and the structure of the intracluster magnetic field.
Also,  neutral pions arising from neutralino annihilation should
give rise to substantial amounts of diffuse gamma-ray emission, up to 
energies of order $M_{\chi}$, that could be tested by
the next generation gamma-ray experiments.
\end{abstract}

\noindent
{\it Subject headings}: cosmology: general -- Dark Matter: neutralino -- 
galaxies: clusters: general -- radio emission: general 
-- radiation mechanisms: non-thermal.

\section{Introduction}

Dark Matter (DM) interaction and annihilation in the halo of 
our galaxy and in other galaxies  have relevant astrophysical implications.
In fact, if DM is constituted by weakly interacting massive particles (WIMP's), 
their annihilation can produce 
direct and indirect signals such as observable fluxes of 
positrons (Silk and Srednicki 1984, Rudaz and Stecker 1988, Ellis et al. 1989, Stecker and Tylka
1989, Kamionkowski and Turner 1991, Baltz and Edsj\" o 1998), 
antiprotons (Chardonnet et al. 1996, Bottino et al. 1998), 
gamma-rays (Bengtsson et al. 1990, Chardonnet et al. 1995) and
neutrinos (Silk and Gondolo 1999) from the Milky Way halo.

Motivated by these results, 
we explore here some specific consequences of the production of positrons and electrons 
from the WIMP interaction in massive DM halos like clusters of galaxies.
The decays of WIMP annihilation products (fermions, bosons, etc.) yield, among other particles, energetic
electrons and positrons up to energies comparable to the WIMP mass (usually of 
the order of tens to hundreds GeV).
It is straightforward to realize that these energetic electrons and positrons (hereafter we will refer
to these particles as electrons because their distinction is not essential for our purposes) 
can emit synchrotron radiation in a DM halo which is filled with a magnetic field at the
level of $\mu$G.
Clusters of galaxies are  the largest bound systems which have indeed both the largest amount 
of DM and extended Intra Cluster (IC)  magnetic field at the level of a 
few $\mu$G.
It is then natural to expect that the WIMP annihilation  products can give origin to 
an extended radio emission.

Extended radio halos and relics are observed, at present, in more than 40 
galaxy clusters 
(see, e.g., Owen et al. 1999, Feretti 1999, Giovannini 1999, Liang et al. 2000).
Cluster radio halos show, generally, a regular morphology, which is similar to the X-ray
morphology, a low surface brightness, and a steep radio spectrum, 
$J(\nu) \sim \nu^{-\alpha_r}$, with $\alpha_r \sim 1 \div 1.5$.
The level of polarization in the radio halo emission is usually $\simlt 10\%$ 
(Feretti 1999, Liang 2000).
The radio halo sizes are typically of $(0.5 \div 1) h_{50}^{-1}$ Mpc, and are centered on
the X-ray emission of the cluster.
Their radio halo luminosity at $1.4$ GHz, $J_{1.4}$, correlates strongly with
the cluster IC gas temperature, $T$, and with its X-ray luminosity, $L_X$, mainly produced by
thermal bremsstrahlung emission (see Colafrancesco 1999, 2000, Liang et al. 2000).
This result points to a strong correlation between the physical state and origin
of the relativistic electrons and the thermal content and distribution of the IC
gas (see Colafrancesco 1999 , 2000 for a more extended discussion).

While we have a definite evidence for the existence of central radio halos in  more than 20 rich
clusters (see Section 6 below), the physical origin of such extended radio halos is still 
a matter of debate.
The main difficulty in explaining the radio halo properties arises from the combination of
their large size ($\sim 1$ Mpc) and the short lifetimes ($\sim 10^8 -10^9$ years)
for synchrotron radiating electrons.
Moreover, also the origin and the evolution of the IC magnetic field is still unclear
(see, e.g.,  Kronberg 1994 for a review).

Successful models for the origin of non-thermal, diffuse radio emission 
from galaxy clusters  have been proposed 
in the context of the possible cosmic ray acceleration mechanisms in 
the intra cluster medium (hereafter ICM).
Synchrotron emission from a population of relativistic electrons 
(Jaffe 1977, Rephaeli 1977, Roland 1981)
accelerated -- or reaccelerated  in situ (Schlikeiser 1987) -- as primary cosmic rays 
(see Sarazin 1999 and references therein), 
or produced {\it in situ} as secondary products of $pp$ collisions 
(Dennison 1980, Vestrand 1982; see also more recently
Blasi and Colafrancesco 1999 and references therein) 
can reproduce the observed spectra of the radio halo emission
from Coma (see Giovannini and Feretti 1996, Deiss et al. 1997 
for an observational review).
In all these models for the origin of radio halos, 
one needs a substantial energy supply to
maintain the brightness of the radio halo for a time comparable with the Hubble time.
The possibility that such energy budget can be supplied to a radio halo from shocks and turbulence
induced by strong cluster mergers (Sarazin 1999, Takizawa 2000)
is suggested by the evidence that several radio halo
clusters (e.g., A1656, A2255, A2163, A2319, A665) are undergoing a merger event.
The shock waves produced by strong mergers could accelerate primary cosmic rays and amplify
the IC magnetic field (see Roettiger 1999) to produce substantial radio powers.
However, there are also some other clusters which still show extended radio halos (e.g., A2256,
A2319, and other clusters found recently by Giovannini et al. 1999) 
but are either in a pre-merger state or even there is no clear merging event occurring.
Moreover,  there are also a few small-size radio halos found in clusters which have cooling flows
(e.g., Perseus, A85, A2218 and maybe A2254).
Therefore, one is bound to assume that: 
{\it i)} the energy released during the very early
phases of the merging process  is sufficient to power  the central radio halo;
or {\it ii)} the electrons are efficiently reaccelerated by the turbulent gas 
motion originated from IC turbulence and/or
the galaxy motion within the cluster (Deiss et al. 1997); 
or {\it iii)} there exists a population of cosmic ray protons accelerated 
by galaxy activity
and/or early cluster mergers which are stored for long times (see, Colafrancesco and
Blasi 1998) in the cluster and produce {\it in situ}
high-energy secondary electrons providing the long-living radio halos
(Blasi and Colafrancesco 1999).
So, while the energy needed for the maintenance of a radio halo can be reasonably accounted for,
the origin of the seed relativistic electrons is not yet fully clear.

In this paper we explore the possibility that the diffuse radio halo emission 
observed in many clusters is produced by high-energy electrons resulting from 
the decay of the annihilation products of DM particles, 
like WIMP's.
Among the  WIMP's candidates, we consider in particular 
the lightest neutralino $\chi$, that is predicted in  supersymmetric extensions
of the Standard Model (Haber and Kane 1985, Jungman et al. 1996). 
The $\chi$ mass is expected to be in the range of tens to
hundreds  GeV. The detailed interaction properties of the neutralino
are determined once its physical composition is fixed. Indeed, neutralinos
are a linear combinations of two Higgsinos (supersymmetric partners of
the Higgs bosons) and two gauginos (supersymmetric partners of the neutral gauge
bosons). 
Both the neutralino mass and its physical composition affect in general 
the rate and the final state composition of the $\chi\chi$ annihilation
process.
Recent analysis of accelerator constraints, combined with cosmological
DM requirements (Ellis et al. 2000), points to:
{\it i}) a lightest neutralino heavier than about 46 GeV;
{\it ii}) a lightest neutralino that is either a {\it pure} gaugino or
a heavily mixed gaugino-Higgsino state, since a predominantly 
Higgsino state cannot provide a substantial component of the dark matter.

According to their physical composition and mass, 
neutralinos can annihilate either into fermion pairs or into vector (or 
Higgs) boson pairs. In general, for a pure-gaugino $\chi$ the annihilation 
into fermion pairs is dominant. Due to the chiral structure
of the initial state, the annihilation cross section inside a halo grows
with the square of the final fermion mass (see, e.g., Turner and Wilczek 1990),
giving larger rates for heavier fermions.
Hence, for pure gauginos, a continuum electron spectrum will arise 
both from the leptonic decays of $b$ and $c$ quarks and $\tau$ leptons,
and from the decays of $\pi$'s originating from the fragmentation
of quark pairs.
On the other hand, when the $\chi$ higgsino component   is not negligible
and $M_{\chi}>M_W$, the annihilation into $WW$ (or $ZZ$) pairs
can become dominant. In this case, the electron spectrum will arise 
partly from the leptonic direct $W$ (or $Z$) decays (giving rise to 
energetic electron of energy $\sim M_{\chi}/2$), and partly from the
hadronic vector boson decays, giving rise to electrons through both the
leptonic heavy-quarks decays and $\pi$'s decays. 

In this paper,  we bypass all the complicancies deriving from
a detailed study of the supersymmetric parameter dependence of the neutralino
annihilation cross sections, by adopting  the following simplified framework, 
that will anyway reproduce the crucial features of the problem at hand:

\noindent
a) we will assume a given value for the neutralino annihilation cross section,
which is both relevant for the DM closure density and inside
the allowed range of the supersymmetric neutralino cross sections for a given 
$M_{\chi}$ and neutralino composition. 
A different value of the cross section can then be straightforwardly rescaled
from our results, as we will discuss in the paper;

\noindent
b) we will study the electron energy spectrum arising from the decays
of the neutralino annihilation products in the two distinct situations
where the dominant annihilation is either into fermions or into vector bosons.
A realistic case will be, of course, close to either these two or to a
linear combination of them.

This simplified scheme will be representative of
a wide range of supersymmetric models that are relevant for the DM problem.   

The plan of the paper is the following.
In Section 2 we discuss the annihilation process of neutralinos in the DM halos of
galaxy clusters, and in Section 3 we give an analytical approximation for the source
spectrum of high-energy electrons produced from the decay of the secondary
products of $\chi \chi$ annihilation. In this Section we also derive the
electron equilibrium spectrum.
In Section 4 we evaluate the radio halo spectrum and its brightness profile from 
synchrotron emission of the relativistic $e^{\pm}$ and
the level of ICS emission from the same population of electrons.
In Section 5 we discuss the
specific cases of the radio halos found in Coma and in the cluster 
1E0657-56 and in Section 6 we show how the
present model is able to reproduce the $J_{1.4}-T$ relation.
We summarize the results of this paper and 
discuss the relevance of the present model for the origin of cluster radio halos 
in Section 7.
The relevant physical quantities are calculated using 
$H_0=50$ km s$^{-1}$  Mpc$^{-1}$  and a flat( $\Omega_0=1$) CDM dominated 
cosmological model.

\section{Neutralino annihilation in galaxy clusters}

Neutralinos decouple from the primeval plasma when they are no longer relativistic.
Their present abundance can be calculated by solving the Boltzmann equation for the evolution of the
density of particle species (see, e.g., Kolb and Turner 1990).

Assuming a spherical, uniform top-hat model for the collapse of a cluster (see, e.g.,  Peebles
1980), the average DM density within a halo which virializes at redshift $z$ is:
\be
\bar {\rho} = \Delta(\Omega_0,z) \rho_b \bigg( 1 - f_g\bigg)
\ee
where the background density is $\rho_b = \Omega_m \rho_c$ and 
$\Delta(\Omega_0,z)$ is the
non-linear density contrast of the virialized halo 
(see, e.g.,  Colafrancesco et al. 1997).
Here $f_g$ is the gas mass fraction.
If we redistribute the total mass, $M={4 \pi \over 3} \bar{\rho} R_p^3$,
found within the radius $R_p= p r_c$ (expressed in terms of the cluster core radius $r_c$), 
according to a density profile $\rho(r) = \rho_0 g(r)$,
we find that the central total density is given by:
\be
\rho_0 = {\bar{\rho} \over 3} {R^3_p \over I(R_p) } ~,
\ee
where $I(R_p) = \int_0^{R_p} dr r^2 g(r)$.
In our phenomenological approach, we consider two cases: 
a constant core model and a central cusp profile.
\newline
The constant core model  with a  density profile:
\be
g(r) = \bigg[ 1 + \bigg({r \over r_c} \bigg)^2  \bigg]^{-\xi/2} ~,
\ee
assumed here to be given from a beta-profile (Cavaliere and Fusco-Femiano 1976) with 
$\xi = 3\beta$ and $\beta = \mu m_p v^2/kT$, gives:
\be
\rho_0 = {\bar{\rho} \over 3} {p^3 \over I(p,\beta) }
\ee
with $p \equiv R_p/r_c$ and $I(p,\beta) = \int_0^p dx x^2 (1+x^2)^{-3\beta/2}$
(here $x\equiv r/r_c$ in terms of the cluster core radius $r_c$).

\newline
The central cusp model described by a profile:
\be
g(r) = \bigg({r \over r_c} \bigg)^{-\eta} \bigg(1+ {r \over r_c} \bigg)^{\eta - \xi} 
\ee
gives:
\be
\rho_0 = {\bar{\rho} \over 3} {p^3 \over I(p,\eta, \xi) } ~,
\ee
where $I(p,\eta, \xi)= \int_0^p dx x^{2-\eta}(1+x)^{\eta-\xi}$.
With $\eta = 1$ and $\xi = 3$, the central cusp model corresponds to the ``universal density profile''
which Navarro, Frenk and White (1997; hereafter NFW) showed to be a good description of cluster DM halos in N-body
simulations of CDM hierarchical clustering.
Note that, in both models the DM density scales, at large radii, as $r^{-\xi}$.

Assuming that the DM density scales like the total cluster density, a general
expression for the central DM number density is:
\be
n_{\chi,0} = {\bar{n} \over 3} {p^3 \over I}
\ee
where the integral $I$ is given in eqs. (4) or (6) for the two DM profiles here considered, and
\be
\bar{n} = 4.21 \cdot 10^{-5} ~{\rm cm}^{-3} ~ \Omega_{\chi} h^2
\bigg[{M_{\chi} \over 100 ~GeV} \bigg]^{-1}
\bigg[{\Delta(\Omega_0,z) \over 400} \bigg] (1-f_g) ~.
\ee
Here we assume that most of the DM consists of neutralinos of mass $M_{\chi}$.

The annihilation rate of neutralinos in a DM halo is
\be
R = n_{\chi}(r) \langle \sigma V \rangle_A ~,
\ee
where $n_{\chi}(r) = n_{\chi,0} g(r)$ is the neutralino number density and
$\langle \sigma V \rangle_A$ is the $\chi \chi$ annihilation 
cross section averaged over a
thermal velocity distribution at freeze-out temperature.
Although, as anticipated, the  $\chi \chi$ annihilation cross section is
a nontrivial function of the mass and physical composition of the neutralino,
to our purpose it suffices to recall that the $\chi$relic density
is approximately given by (Jungman et al. 1996):
\be
\Omega_{\chi}h^2 \simeq \frac{3 \times 10^{-27}~{\rm cm}^{3}~{\rm s}^{-1}}
                  {\langle \sigma V \rangle_A }~~.
\ee
Hence, for the $\chi \chi$  annihilation, we will
assume a total $\chi$ cross section  of
\be
\langle \sigma V \rangle_A \approx 10^{-26} ~{\rm cm}^3  ~{\rm s}^{-1} ~~
\ee
to be consistent with the value $\Omega_m \sim 0.3$ derived from clusters of galaxies
and large scale structure constraints (see, e.g., Bahcall 1999).
Detailed studies of the relic neutralino annihilation (Edsj\"o 1997)
show that  the above value is well inside the allowed range predicted
in supersymmetric theories for a wide choice of masses and 
physical compositions of neutralinos that can be relevant as cold 
DM candidates.
Enhancing  the $\chi$ annihilation rate will have on our results
the simple effect of rescaling the final electron spectra by the same 
enhancement factor

For values of $h$ in the range
$0.5 \simlt h \simlt 1$, we can infer that for a flat, vacuum dominated  CDM 
universe with 
$\Omega_{\Lambda}$ in the range $0.5 \div 0.7$ and $\Omega_{\chi} \sim 0.3\div 0.5$, 
the quantity $ \Omega_{\chi} h^2$ takes values in the range $\sim  0.075 \div 0.5$, which would fix the annihilation 
cross section to within a factor less than ten.

\section{Production and equilibrium electron spectra}

\subsection{The source spectrum}

Neutralinos which annihilate in a DM halo produce quarks, leptons,
vector bosons and Higgs bosons, depending on 
their mass and physical composition (see, e.g., Edsj\"o 1997).
Following the discussion in Section 1, monochromatic electrons 
(with energy  about $M_{\chi}$), 
coming from the direct channel $\chi\chi \to ee$, 
are in general much suppressed (Turner and Wilczek 1990).  
Electrons are then produced  from the decay 
of the final heavy  fermions and bosons.

The different composition of the $\chi\chi$ annihilation final state 
will in general affect the form of the final electron spectrum.
To our purpose, it will be sufficient to consider 
two somewhat extreme cases:
1)  a pure-gaugino annihilation,  which yields mainly fermion 
pair direct production $\chi\chi \to ff$, with cross sections scaling
as $M_f^2$ for different fermions (Turner and Wilczek 1990);
2) a mixed gaugino-higgsino state in the case of a dominant annihilation
into $W$ (and $Z$) vector bosons, $\chi\chi \to WW (ZZ)$.
A real situation will be mostly reproduced by either of the above two cases, 
or by a linear combination of the two.

The positron spectrum arising from the $\chi \chi$ annihilation has 
been derived by various
authors (Silk and Srednicki 1984, Rudaz and Stecker 1988, Ellis et al. 1989, 
Stecker and Tylka 1989, Turner and Wilczek 1990, Kamionkowski and Turner 1991, 
Baltz and Edsj\" o 1998).
Here, we will adopt  the  approaches by Rudaz and Stecker 
(1988; hereafter RS) 
and by Kamionkowski and Turner (1991; hereafter KT),
that gave the analytical approximations of 
the positron source functions for models in which neutralinos
annihilate mainly into fermions and vector bosons, respectively.
Here we report the results of our calculations while
the details of these two approaches are presented in the Appendix.

The total source spectra in the two cases  considered are shown 
in Fig.1  for $M_{\chi}=100$GeV.
The heavy solid curve is the total spectrum for the model in which fermions 
dominate the annihilation.
We also show the different contributions to 
the total source spectra from fermions, i.e.
the source spectrum for first-generation prompt electrons (P1),
second-generation prompt electrons (P2) and secondary electrons produced in 
the decay of charged pions ($\pi$).
The total source spectrum  is rather smooth and, when approximated by a single
power-law, it has an overall slope $Q_e \sim E^{-1.9}$ in the interesting energy range
$0.02 \leq E/M_{\chi} \leq 0.7$.

The source spectra for the model in which gauge bosons 
dominate the annihilation is given by the light solid curve.
For this case we show the contributions from the decay 
$W^{\pm} \to \tau^{\pm} \to e^{\pm}; ~ W^{\pm} \to c \to e^{\pm}$ (labelled as KT$_1$)
and from the decays of charged pions produced in hadronic decays of the $W^{\pm}$
(labelled with KT$_{\pi}$).
Note that we make this analysis restricting only to the $WW$ channel, since the $ZZ$ contribution
gives rise to a qualitatively similar spectrum.

Two prominent spectral 
features (bumps) around $E \sim M_{\chi}/2$ (arising from the direct decay
$W\to e\nu$) and $E \sim M_{\chi}/20$ are 
shown by the model in which gauge
bosons dominate the annihilation, in comparison with the smoother spectral 
shape of the fermion dominated  annihilation.
These spectral features  are more prominent in Fig.2 where we plot the quantity $E^3
Q_e(E,r)$.
From a closer inspection of Figs.1 and 2, one can see that the main difference 
between the two models stands in the height and width of the bumps at
energies below $M_{\chi}/10$, where the pion-produced electrons 
dominate the source spectrum, and in the high-energy tail at $E \simlt 0.7 M_{\chi}$
where the P1 contribution dominates.
In both models, electrons produced by tertiary decays of gauge 
bosons are neglected because their
contribution  is subdominant with respect to the pion decay distribution.

In the calculation of the radio halo emission (see Section 4 below) the source spectra
of eq.(A1) and A(18) will be multiplied by a factor 2 to take into account the
contribution of both electrons and positrons.

\subsection{The electron equilibrium spectrum}

The time evolution of the electron spectrum is given by the transport equation:
\be
{\partial n_e (E,r)\over \partial t} - {\partial  \over \partial E} 
\bigg[ n_e (E,r) b(E)\bigg] = Q_e(E,r)
\ee
where $n_e(E,r)$ is the equilibrium spectrum at distance $r$ from the cluster
center for the electrons with energy $E$.
The source electron spectrum rapidly reaches its  equilibrium configuration mainly
due to synchrotron and Inverse Compton Scattering (hereafter ICS) losses at energies $E \simgt 150$
MeV and to Coulomb losses at smaller energies (Blasi and Colafrancesco 1999).
Since these energy losses are efficient in the ICM and DM annihilation continuously
refills the electron spectrum, the population of high-energy electrons
can be described by a stationary transport equation ($\partial n_e / \partial t \approx 0$) 
\be
 - {\partial  \over \partial E} \bigg[ n_e (E,r) b(E)\bigg] = Q_e(E,r)
\ee
from which the equilibrium spectrum can be calculated.
Here, the function $b(E)$ gives the energy loss per unit time at energy $E$
\be
b_e(E) = \left(\frac{dE}{dt}\right)_{ICS} +
\left(\frac{dE}{dt}\right)_{syn} +
\left(\frac{dE}{dt}\right)_{Coul}
= b_0(B_{\mu}) E^2 + b_{Coul} ~,
\ee
where
$b_0(B_{\mu})= (2.5\cdot 10^{-17} + 2.54 \cdot 10^{-18} B_{\mu}^2)$
and $b_{Coul}= 7\times 10^{-16} n(r)$
(if $b_e$ is given in units of GeV/s). 
In the expression for $b_{Coul}$, the IC gas density, $n(r)$, 
is given in units of $cm^{-3}$.

The equilibrium spectrum calculated combining the source spectrum $Q_e(E,r)$ and the energy losses
$b(E)$ is shown in Fig.3 for the two cases here considered.
Equilibrium spectra are evaluated separately for the different components of the source spectra and then
summed together.
This spectrum is evaluated within the cluster core 
for a constant core density (see eq.2) with a reference value of
the central density $n_{\chi,0}=1 ~cm^{-3}$ and $\langle \sigma V \rangle_A=10^{-26}$
cm$^3$ s$^{-1}$.
This equilibrium spectrum can be easily calculated for other cluster configurations by scaling the
parameters of the cluster density profile.
The equilibrium spectrum for the case in which neutralino annihilation is dominated by fermions, is
smooth enough that it can be reasonably approximated by a power-law, $n_e(E) \sim E^{-p}$,
down to $E \sim M_{\chi}/30$ with a slope $p \sim 2.9$.
We will make use of this convenient approximation in the discussion of our results.

\subsection{Comparison between approximated and Monte Carlo results}
The source spectra in Fig. 1 arise from an analytic approximation (see the Appendix) 
of the exact shape of the electron spectrum that tries to cope with 
the details of the quarks and 
leptons decays and of the hadronization of the decay products. 
Detailed \el spectra can also  be obtained by using state-of-the-art Monte 
Carlo simulations, 
although the analytical approximations used here are able to 
recover the relevant aspects of more detailed studies.

There are, however,  some differences among the equilibrium spectra
obtained from different studies using Monte Carlo techniques.
We can compare here our analytical approximations with
the results of the Monte Carlo simulations run by Baltz and Edsj\" o (1998), 
and Golubkov and Koplinoch (1998).
Note that the latter study considers massive neutrinos annihilation, that, as far as
the electron spectra shapes are concerned, should be equivalent to 
the neutralino annihilation
at fixed $M_{\chi}=m_{\nu}$, and fixed $ff$/$WW$ annihilation dominance. 
One can note  that for the case we consider here, i.e. $M_{\chi} = 100$ GeV, 
the total equilibrium electron
spectrum from the simulations of Golubkov and Konoplich (1998) agrees quite 
well with the spectrum shown
in Fig.3 for the case in which neutralinos annihilate mainly into fermions.
On the other hand,
the case in which the annihilation is dominated by vector bosons (KT) is quite 
well reproduced by the
simulations in Baltz and Edsj\" o (1998), although there a slightly different
value of  $M_{\chi}$ (i.e., $M_{\chi} \sim 130$ GeV) is assumed.

However, beyond the limits of the present discussion, it is worth noticing that 
there are still differences in the electron spectra so far published
by different authors using Monte Carlo techniques.
In view of this,
we conservatively assume in our calculations a reasonable uncertainty of a 
factor $\sim 2$  in the  amplitude of the electron spectrum, and of 
$\sim 10 \%$ in its slope.
We will consider more detailed predictions 
derived from available Monte Carlo simulations in a forthcoming paper.

\section{The diffuse radio halo emission}

The calculation of the radio emissivity per unit volume is
performed here in the simplified assumption that electrons with energy $E$
radiate at a fixed frequency given by \cite{longair}:
\begin{equation}
\nu \approx 3.7 ~{\rm MHz}~ B_{\mu} \bigg( {E \over {\rm GeV}} \bigg)^2 ~.
\label{eq:freq}
\end{equation}
This approximation introduces negligible errors in the final result and
has the advantage of making it of immediate physical interpretation.

The radio emissivity at frequency $\nu$ and at distance $r$ from
the cluster center can be calculated as
\begin{equation}
j(\nu,r)  = n_e(E,r) \left(\frac{dE}{dt}\right)_{syn}~
\frac{dE}{d\nu}~,
\label{eq:j2}
\end{equation}
where $(dE/dt)_{syn}$ is given in eq.(14).
The radio halo luminosity 
is obtained by integration of $j(\nu,r)$ over the cluster
volume yielding:
\be
J(\nu) = 4 \pi \int_0^{R_{halo}} dr ~ r^2 j(\nu,r) ~,
\ee
where $R_{halo}$ is the size of the cluster radio halo.
The observed radio halo flux is then
$$
F(\nu) = {J(\nu) \over 4 \pi D_L^2}
$$
where $D_L(z,\Omega_0)$ is the cluster luminosity distance (see, e.g.,  Weinberg 1972).
For the sake of illustration, we give the explicit form of $J(\nu)$ in the case of a
power-law source spectrum, $Q(E,r) = Q_0 ( E/E_0)^{-s}$, where
$Q_0= n_{\chi}^2(r) \langle \sigma V \rangle_A (k M_{\chi})^{-1}$ and $E_0$ is a
normalization energy, say $1$ GeV.
Using eqs. (13), (16) and (17), we can write
\be
J(\nu) = {4 \pi n_{\chi,0}^2 \langle \sigma V \rangle_A (k M_{\chi})^{-1} \over 2 (s-1)}
\bigg( {E_{\ast} \over E_0}\bigg)^{-s+2} E_0 {\cal I} 
{b_{0,syn}(B_{\mu}) \over b_{0}(B_{\mu})} B_{\mu}^{\alpha_r -1} \nu^{-\alpha_r}
\ee
where $\alpha_r=s/2$,  $E_{\ast}=16.44$ GeV, 
${\cal I} = \int_0^{R_{halo}} dr r^2 g^2(r)$,
$b_0(B_{\mu})= (2.5\cdot 10^{-17} + 2.54 \cdot 10^{-18} B_{\mu}^2$ and
$b_{0,syn}(B_{\mu})= 2.54 \cdot 10^{-18} B_{\mu}^2$.
For $B_{\mu} \simlt 3.14$, the radio luminosity scales in eq.(18) as
$J(\nu) \propto n_{\chi,0}^2 \langle \sigma V \rangle_A (k M_{\chi})^{-1} 
B_{\mu}^{\alpha_r +1} \nu^{-\alpha_r}$.

In the calculation of the radio halo luminosity 
we consider both the case of a uniform IC magnetic field and a case 
in which the magnetic field has a radial profile
which declines with the distance from the cluster center, as indicated by recent 
numerical simulations (see, e.g., Friaca and Goncalves 1998).
Specifically, we considered a magnetic field density profile
\be
B = B_0 \bigg[ 1 + \bigg({r \over r_{c,B}} \bigg)^2  \bigg]^{-w/2}
\ee
with values $w \sim 0.5\div 1$.
Here, for simplicity,  we assume that the spatial structure of the magnetic field is given by a
constant core profile with lenght scale $r_{c,B}$.

\subsection{The radio halo spectrum and spatial distribution}
The total radio halo flux predicted from our model
has generally a steep spectrum with average slope $\alpha_r \sim 1.2 - 1.8$,
depending on the neutralino mass and composition and on the shape of the DM  density and 
magnetic field profiles.
We will discuss more specifically this point in Section 5 below. Here we want to
emphasize that
for a fixed value of the magnetic field $B$,
the neutralino mass fixes the maximum frequency at which we can observe synchrotron emission from 
the radio halo, 
\be
\nu_{max} \sim 3.7 ~{\rm MHz} ~B_{\mu} \bigg({ k M_{\chi} \over GeV} \bigg)^2
\ee
For $M_{\chi} = 100$ GeV and $k \approx 0.7$ (see Appendix), 
the maximum frequency at which the radio halo spectrum is observable is 
$\nu_{max} \approx 18.1$ GHz $\times B_{\mu}$. 
From this fact, we can use the available data on the radio halos of Coma and 1E0657-56 
(see Figures 3 and 4) to set a lower limit to the neutralino mass.
Specifically, the maximum frequency, $\nu_{max,obs}$, at which the radio halo spectrum is observed sets
a lower limit
\be 
M_{\chi} \geq {16.44 \over k} ~{\rm GeV}~\bigg({\nu_{max,obs} \over {\rm GHz}} \bigg)^{1/2} B^{-1/2}_{\mu}
\ee
which gives $M_{\chi} \geq 74.3$ GeV $B^{-1/2}_{\mu}$, for $\nu_{max,obs} = 10$ GHz.
Furthermore, the present theory predicts a sharp cutoff in the radio halo spectrum 
at $\nu_{max}$ given by eq.(20);
so, high frequency observations can both test the neutralino model and measure the quantity 
$B_{\mu} (k M_{\chi})^2$ from the possible detection of the high-$\nu$ sharp cutoff in
the radio halo spectrum.

The brightness profile of the radio halo is obtained by integration of the emissivity,
$j(\nu, r)$, along the line of sight. 
It scales, for a power-law radio halo spectrum, as
\be
S_{radio} \propto \int d \ell ~n^2(r) \cdot B^{1+\alpha_r}(r)
\ee
and declines at large radii  as $r^{-[\xi+w(1+\alpha_r)] + 1/2}$.
The radio halo brightness profile resembles, for both the DM profiles here considered, the 
X-ray brightness profile of the cluster which scales as
\be
S_{X-ray} \propto \int d \ell ~n^2(r) \cdot T^{1/2}(r) ~.
\ee
In particular, for a uniform magnetic field, $B =$ const, 
and for a isothermal cluster, $T=$ const., the brightness $S_{radio}$ behaves
exactly as the X-ray brightness $S_{X-ray}$ at all radii.
Small deviations from this behaviour can be attributed to radial variations in 
$B(r)$ and/or in $T(r)$.

The radio halo brightness profile depends on the assumed DM profile (see eqs.19 and 22).
Specifically, at small radii the constant core model still gives a radio brightness
profile that resembles the X-ray emission  profile 
while the central cusp model tends to give a radio brightness which is 
more peaked toward the center.
Thus, the model predicts a strong radio emissivity at the cluster center under
the assumption of a central cusp (or NFW) profile.
However, neutralino annihilation can also cause a softening of the central DM cusp 
due to the high interaction rate of neutralinos in the central region of the clusters with respect to the 
outer parts (see Kaplinghat, Knox and Turner 2000). This countereffect 
weakens the aforementioned problem.

\subsection{Inverse Compton Scattering emission}
The relativistic electrons which are responsible for the radio halo emission
also emit X-rays and UV photons through Inverse Compton Scattering (ICS) 
off the Cosmic Microwave Background photons. 
As in the case of the synchrotron emission, also for ICS
we can adopt the approximation that electrons radiate at a single energy,
given by
\begin{equation}
E_X=2.7~keV ~E^2(GeV).
\label{eq:ex}
\end{equation}
Electrons with energy in excess of a few GeV radiate in the
hard X-ray range, while electrons with energy smaller than $\sim 400$
MeV produce soft X-rays and UV photons.
The non-thermal X-ray/UV emissivity at distance $r$ from the
cluster center  is evaluated as
\begin{equation}
\phi_X(E_X,r) = n_e(E,r) \left(\frac{dE}{dt}\right)_{ICS}~
\frac{dE}{dE_X} ~.
\label{eq:phi2}
\end{equation}
\noindent
In complete analogy with the case of the radio emission,
the integrated non-thermal X-ray luminosity, $\Phi_X(E_X)$ is
\begin{equation}
\Phi(E_X,r) =
\int_0^{R_{halo}} dr ~4 \pi r^2 \phi_X(E_X,r) ~.
\label{eq:phi2_1}
\end{equation}
The predicted X-ray fluxes from ICS emission due to the electrons produced in $\chi
\chi$ annihilation
are lower than those required to explain the 
hard X-ray excesses currently observed in Coma, A2256 and A2199, 
and do not contribute substantially to the EUV and
hard X-ray emission excesses of these clusters.

\section{Application to observed radio-halo clusters: Coma and 1E0657-56}
Our theory for the origin of radio halos in clusters can reproduce successfully the
spectra of the radio halo emission observed from Coma (Giovannini et al. 1993, Deiss et
al. 1997) 
and 1E0657-56 (Liang et al. 2000). Moreover, these data in turn can set interesting
constraints to the mass and composition of the neutralino.

For any cluster, we choose a DM central density, $n_{\chi,0}$, 
which is given by eq. (7) evaluated with the parameters appropriate to each cluster
and, in the case of 1E0657-56, with parameters approriate to each reagion in which the
radio halo spectrum has been observed.


In the case of Coma we choose the following parameters:
$r_c = 0.4 h^{-1}_{50}$ Mpc, $\beta=0.76$, $n_{\chi,0}=3 \cdot 10^{-3}$ cm$^{-3}$.
The radio halo spectrum has been integrated out to a radius 
$R_{halo} = 1.3 h^{-1}_{50}$ Mpc (see, e.g., Giovannini et al. 1993).
In the model in which neutralino annihilation is dominated by fermions ($\chi \chi \to
f f $),  
a uniform magnetic field $B_{uniform} \approx 1.3 ~\mu$G  is needed to fit the spectrum
under the assumption of a constant core profile.
In the case of a declining magnetic field $B= B_0 [1+(r/r_{c,B})^2]^{-0.7}$, 
a value $B_0=8$ $\mu$G fits better the data, under the same model assumptions.
The resulting spectra are shown in Fig.4.
In this figure we also show the spectra derived from the power-law approximation of
the equilibrium spectrum shown in Fig.3.
A radially decreasing magnetic field yields a slightly steeper spectral slope, as it
should be expected from eq.(17).

Under the same assumptions of a constant core profile and radially decreasing 
magnetic field, the model in which neutralino annihilation is dominated by gauge bosons 
($\chi \chi \to WW $) does not fit the Coma radio halo (see Fig.4).
Within the limits of the analytical approximations used in this paper, 
the Coma data exclude this model at more than $3$ standard deviations (see Fig.5).
In this figure we also show the range of uncertainties in the final radio halo spectra
calculated considering an uncertainty of a factor $\pm 2$ in the overall
normalization of the source spectra and $\pm 10 \%$ 
uncertainty on their slopes, for the two extreme annihilation model here
considered.


In the case of the distant cluster 1E0657-56 at $z=0.269$, we choose different
parameters according to the two different regions in which the spectrum has beem
measured (we use the core radii and $\beta$ parameters given in Liang et al. 2000).
We assume that neutralino annihilation is dominated by fermions ($\chi \chi \to f f$)
and consider a constant core DM density profile.
For the outer region we choose
$r_c = 0.38 h^{-1}_{50}$ Mpc, $\beta=0.7$, $n_{\chi,0}=9 \cdot 10^{-3}$ cm$^{-3}$.
The radio halo spectrum has been integrated out to a radius 
$R_{halo} = 2 h^{-1}_{50}$ Mpc (Liang et al. 2000).
A uniform magnetic field $B_{uniform} \approx 2. ~\mu$G is needed to fit the spectrum
of the outer region.
In the case of a declining magnetic field $B= B_0 [1+(r/r_{c,B})^2]^{-0.5}$, a value 
$B_0=100$ $\mu$G can fit the data.
For the inner  region we choose
$r_c = 0.08 h^{-1}_{50}$ Mpc, $\beta=0.49$ and the same central density 
$n_{\chi,0}= 9 \cdot 10^{-3}$ cm$^{-3}$.
The radio halo spectrum has been integrated out to a radius 
$R_{halo} = 0.6 h^{-1}_{50}$ Mpc.
A uniform magnetic field $B \approx  9~\mu$G is needed to fit the spectrum of this
region.
In the case of a declining magnetic field $B= B_0 [1+(r/r_{c,B})^2]^{-0.5}$, a value 
$B_0=100$ $\mu$G fits the data.
The results are shown in Fig.7.
As in the case of Coma, the extreme case of neutralino annihilation dominated by gauge
bosons ($\chi \chi \to W W$) is inconsistent with the radio halo spectrum of this
cluster.


Note that the radio-halo spectra shown in Figures 4-7 
are calculated assuming a constant core
density profile for the DM component.
A central cusp model gives a higher central density and increases the
$\chi$ annihilation rate in the central region of the cluster which in turn produces
more high-energy electrons per unit volume.
This increase in the electron density must be compensated by a decrease in the
magnetic field value needed to fit the radio halo spectrum.
In fact, assuming a NFW density profile for Coma, a uniform magnetic field 
$B_{uniform} = 0.45 \mu$G is required to fit the data.
Alternatively, a central value of the magnetic field, $B_0= 1.4 \mu$G is able to
fit the spectrum assuming a $B$ radial profile as given by eq. (19).
We show the difference between the spectra obtained either 
with a beta-profile or with a NFW profile in Figure 8.

\section{The $J_{1.4}-T$ relation for radio-halo clusters}
Another interesting property of radio halo clusters is the steep correlation 
existing between the monochromatic radio halo luminosity 
observed at $1.4$ GHz, $J_{1.4}$, and the IC gas temperature:
$J_{1.4} \sim T^q $ with $q=6.25$ (range $q= 4.1 \div 12.5$ at $90 \%$ confidence 
level) found by Colafrancesco (1999, 2000) and confirmed by Liang et al. (2000).
Radio luminosity data are taken from Feretti (1999), Giovannini et al. (1999), Liang et
al. (1999) and Owen et al. (1999).
Here, we use the temperature data from Arnaud and Evrard (1999), bur a similar result 
obtains using temperature data from Wu, Xue and Fang (1999) or David et al. (1993).
Such a steep correlation (see Fig.9) can be reproduced in models of radio halos 
powered by
acceleration of cosmic rays triggered either by strong merging events or by enhanced 
galaxy activity, but requires specific conditions for the structure of the magnetic 
field  (see Colafrancesco 1999, 2000 for a discussion).

In the model we present here the radio halo luminosity (see eqs.17-18) scales as
\be
J(\nu) \propto n_{\chi,0}^2 r_c^3 \langle \sigma V \rangle_A B^{1 + \alpha_r} \nu^{-\alpha_r}
\ee
where $\alpha_r = d ln J(\nu) / d ln \nu$ is the effective slope of the radio halo spectrum.
Thus, the monochromatic radio halo luminosity correlates naturally with the cluster X-ray bremsstrahlung luminosity, 
$L_X \propto n_0^2 f_g^2 r_c^3 T^{1/2}$ (here $n_0$ is the central IC gas density and
$f_g= M_{gas}/M$ is the cluster baryon fraction), to give
\be
J_{1.4} \propto L_X f_g^{-2} T^{-1/2} \langle \sigma V \rangle_A B^{1 + \alpha_r} ~.
\ee
Using the observed correlation, $L_X \sim T^{b}$ with $b \sim 3$ (see, e.g., Arnaud
and Evrard 1999, Wu et al.  1999, David et al. 1993)
and assuming that $f_g= const$, as indicated by the available cluster data 
(Jones and Forman 1999, Mohr 2000), we derive 
\be
J_{1.4} \sim T^{(4b-1)/2}  \langle \sigma V \rangle_A B^{1 + \alpha_r}
\ee
Under the condition -- assumed in this paper -- of hydrostatic equilibrium of the 
cluster material with the overall potential wells, the relation
$V \sim T^{1/2}$ holds, and we finally obtain:
\be
J_{1.4} \sim T^b  B^{1 + \alpha_r} ~.
\ee
If every cluster has the same value of the magnetic field (universality condition), 
a correlation $J_{1.4} \sim T^{3}$ obtains.
This is too flat compared to the observed correlation.
However, the universality condition $B=$ const is rather unlikely.
In fact, we expect that $B$ increases
with increasing $T$ either under the condition of energy equipartition between the 
IC gas and the magnetic field (which yields $B \sim T^{1/2}$) or 
even under the condition of a $B$ field which is frozen-in 
the ICM (i.e., $B \sim n^{\gamma}$ with $\gamma \sim 2/3$, 
which yields $B \sim T^{(b-2)\gamma}$).

In conclusion, it is reasonable to expect a quite steep correlation, 
$J_{1.4} \sim T^{4.15 \pm 0.25}$ or $J_{1.4} \sim T^{4.40 \pm 0.26}$, 
for an equipartition or frozen-in magnetic field.
Such a correlation is consistent (within the uncertainties) 
with the best fit result given above and 
can reproduce the overall
distribution of clusters in the $J_{1.4} - T$ plane.
Moreover, since $T \propto (1+z) \Delta^{1/3}(\Omega_0,z)$ in CDM models of 
structure formation (see, e.g., Colafrancesco et al. 1997), 
the radio halo power is expected to evolve strongly with  redshift 
$J_{1.4} \sim [(1+z) \Delta^{1/3}(\Omega_0,z)]^q$.
Thus, already at $z\sim 0.2$ its normalization 
increases by a factor $\sim 2$ (we use here $q = 4$ as a reference value).
The $J_{1.4} - T$ correlation predicted in the present
model at $z = 0 $ and at $z = 0.25$ is shown in Fig.9.
The large steepness $q \approx 6.25$ deduced by a formal power-law fit to the data
can be due to a superposition of the different correlations of radio halo
clusters observed in the range $z \approx 0 \div 0.3$.

\section{Discussion and conclusions}
In this paper we have shown that the basic properties 
(i.e., the spectrum, the surface brightness profile and  the $J_{1.4} - T $ 
correlation) of radio halos in galaxy clusters can be
reasonably fitted by a model in which the high-energy electrons,  
giving rise to the synchrotron emission, arise from  the decay of secondary products of the  neutralino 
annihilation in the DM halos of galaxy clusters.
The slope and the frequency extension of the spectrum can set interesting 
constraints on the neutralino mass and composition.
In fact, the highest observed frequency of the radio halo spectrum set a lower 
limit to the neutralino mass: 
$$
M_{\chi} \geq 16.44 ~{\rm GeV} ~k^{-1}\bigg[ \bigg( {\nu_{max,obs} \over GHz} \bigg) {1 \over B_{\mu}}
\bigg]^{1/2}~.
$$ 
In particular, the highest frequency  
($\nu \sim 4.85$ GHz) at which the Coma spectrum has been observed 
requires $M_{\chi} \geq 54.6 ~{\rm GeV}~ B_{\mu}^{-1/2}$. 
A more stringent limit, $M_{\chi} \geq 70.5 ~{\rm GeV}~B_{\mu}^{-1/2}$, 
is obtained from the cluster 1E0657-56.
Radio halo observations at frequencies larger than $5 \div 10$ GHz 
are not yet available, but are important to test the present model.
In fact, the detection of a high frequency cutoff, $\nu_{cut}$, in the radio halo spectra 
gives an upper limit on the neutralino mass. 
The high-$\nu$ cutoff predicted in our model, has the same value, 
$\nu_{cut} = 3.7 ~{\rm MHz} B_{\mu} [k(M_{\chi}/GeV)]^2$, for all radio halos and 
depends only on the neutralino mass and the IC magnetic field.
For a fixed neutralino mass and composition, the observations of this high-$\nu$ cutoff can yield a
direct estimate of the cluster magnetic field.

In the same framework,
the slope of the radio halo spectrum gives an indication on the 
neutralino physical composition.
The radio halo spectrum of the two clusters considered here are well 
fitted by a 
model in which the neutralinos behave like pure gauginos, and annihilate 
mainly into fermions.
On the other hand, when the neutralino annihilation is dominated by vector 
bosons (implying a non-negligible higgsino component), 
the electron source spectrum is too steep and shows two unobserved bumps at low and high
energies (see Figs.1 and 2). 
This feature is also evident in the final radio halo spectrum (see Fig.5).
Indeed, Figs. 5 and 6 show how far the neutralino model dominated by vector bosons is from reproducing the data of the
radio halo spectrum of Coma and 1E0657-56.

After setting the neutralino mass and composition, the slope of the
radio halo spectrum still depends on the DM density profile and on the 
magnetic-field radial profile, that influence the annihilation rate and the
synchrotron emission power, respectively.
The available data on Coma indicate (in the case of a constant core density 
profile) that a radially decreasing magnetic field is favoured to fit
the correct steepness of the spectrum at high frequencies.
This constraint is  tighter for Coma, where an upper limit of 
$52$ mJy at $\nu = 4.85$ GHz has been measured (see, e.g., Deiss et al. 1997), 
than for 1E0657-56, where a rather constant slope is 
observed up to $\nu \sim 10$ GHz. 
For a constant core density profile, the spectra of Coma and 1E0657-56 are 
consistent either with  uniform IC magnetic fields $B_{uniform} \sim 1.3  \div  2~ \mu G$ or 
central values  $B_0 \sim 8 \div 100~ \mu$G for a magnetic field profile 
$B(r) \sim r^{-1}$ at large distances from the cluster center.
Using a central cusp (NFW) density profile increases slightly the slopes of 
the radio halo spectrum, reducing consequently the need for a magnetic field 
strongly decreasing with the distance from the cluster center.
The central cusp profile requires also IC magnetic field amplitudes 
lower by a factor $\sim 3 \div 5$ to fit the radio halo spectrum.

The present model also predicts a mild steepening of the radio halo spectrum with increasing
distance from the cluster center. This effect is  mainly due both to the
lower DM densities (which decrease the neutralino annihilation rate) and to a 
decrease of  the IC
magnetic field (which reduces the emitted powers via synchrotron emission)
at large distances from the cluster center. 
The model presented here can fit reasonably well the radio halo spectra of the cluster 
1E0657-56 in the inner and outer regions at which it has been measured by 
Liang et al. (2000).
It also gives values for the radio spectral index of Coma 
$\alpha_r \sim 0.9 \div 1.6$, from the cluster center out to $\sim 1$ Mpc. These values are 
consistent (within the given uncertainties) with the increase of the
radio spectral index of Coma  given by Giovannini et al. (1993).

It is also noticeable that the spatial extension of the radio halo emission 
which is predicted in the present model is very similar to that of the X-ray 
emission from the IC gas, a property which has been emphasized in several observational works (see, 
e.g., Feretti 1999, Liang et al. 2000).
Changes of radio halo/X-ray surface brightness ratio, $S_{radio}/S_{X-ray}$, with
radius can be accounted for by radial variations of the IC magnetic field and/or the
IC gas temperature.

The analysis presented in this paper is made for a value of the neutralino
annihilation cross-sections $\langle \sigma V \rangle_A \approx 10^{-26} ~cm^3 s^{-1}$.
This value is well inside the allowed range for a neutralino relevant
as a DM candidate, and can be rescaled
in a straightforward way from our results,
for different assumptions on $\langle \sigma V \rangle_A$.
Because the radio halo spectrum power (see eqs.17-18) scales linearly with 
$\langle \sigma V \rangle_A n_{\chi}^2 B^{1+\alpha_r}$ 
(here $\alpha_r= dlnF(\nu) / dln \nu$ is the effective spectral slope),
higher (lower) values of the annihilation cross section imply lower (higher) 
DM densities, and require a lower (higher) IC magnetic field.

Another interesting aspect of the present model for the origin of cluster 
radio halos 
is that it can reproduce fairly well the observed correlation between the
monochromatic radio halo power, $J_{1.4}$, and the IC gas temperature $T$.
The steep $J_{1.4}-T$ relation shown by the available radio-halo clusters can 
be reproduced in our
model as a superposition of evolutionary effects of the
correlation $J_{1.4} \sim T^q$,
with $q \sim 4.2  \div 4.4$, which results from the dependence of the electron
spectrum,
$n_e(E,r) \propto n_{\chi}^2(r) \langle \sigma V \rangle_A$, from 
the DM density, $n_{\chi}(r)$, and annihilation cross-section,
$\langle \sigma V \rangle_A$,  
with the further assumption of an energy 
equipartition  between the IC gas and the IC magnetic field.

The ICS emission from the same population of relativistic electrons produces 
unavoidably also fluxes of UV and X-ray emission that, however, 
are not so intense to reproduce the emission excesses in the EUV 
(Lieu et al. 1996) and in the hard X-rays 
(Fusco-Femiano et al. 1999, Rephaeli, Gruber and Blanco 1999) observed in Coma.
This fact should be not considered as a problem for the model we have worked out 
here since other 
alternative explanations have 
been proposed to fit the EUV and hard X-ray
emission excesses in the framework of thermal (see Antonuccio-Delogu et al. 2000) 
and/or suprathermal phenomena
(Ensslin et al. 1999-2000, Dogiel 1999, Sarazin and Kempner 1999, Blasi et al. 
2000), respectively .

Another remarkable feature of the present model, is that 
neutralino annihilation can also give rise to gamma rays with
continuum fluxes  which are overwhelmingly due to $\pi^0 \to \gamma + \gamma$ decays.
The continuum gamma-ray spectrum is given by:
\be
f_{\gamma}(E_{\gamma}) = 2 \int_{E_{\ell}(E_{\gamma})}^{M_{\chi}} 
dE_{\pi} (E_{\pi}^2 - m^2_{\pi})^{-1/2}
\zeta_{\pi} f(E_{\pi})
\ee
(see,e.g., Rudaz and Stecker 1988) 
where the quantity $\zeta_{\pi} f(E_{\pi})$ is derived in equation (A12) of the 
Appendix.
We recall here that $E_{\pi}$ is the pion energy, 
$E_{\ell}= E_{\gamma}+ m^2_{\pi}/4 E_{\gamma}$  
and $E_{\gamma}$ is the energy of the gamma-ray photon.
The resulting gamma-ray luminosity is
\be
L_{\gamma} = 4 \pi n_{\chi,0}^2 \langle \sigma V \rangle_A \int_0^R dr r^2 g^2(r) 
f_{\gamma}(E_{\gamma})
\ee
and yields, for Coma, a gamma-ray flux $F_{\gamma}(>100 MeV) = 3 \cdot10^{-9}
~cm^{-2} s^{-1}$ (for $M_{\chi}=100$ GeV; see Fig.10), which stays below the 
EGRET upper limit for Coma, $F^{Coma}_{\gamma}(>100 MeV) = 4 \cdot10^{-8}
~cm^{-2} s^{-1}$ (Sreekumar et al. 1996).

Note that a correlation 
$L_{\gamma} \propto L_X^{(2b-1)/2b}$ (here $b$ is the 
exponent of the X-ray luminosity -- temperature correlation, $L_X \sim T^b$) 
is expected in this model with $f_g=$ const, at variance with the correlation 
$L_{\gamma} \sim L_X^{1/4}$ expected in models of cosmic ray ($pp$) interaction in the
ICM, (Colafrancesco and Blasi 1998).
The different slope of the $L_{\gamma} -L_X$ relation will provide a way to disentangle between these
two mechanisms for the production of gamma rays in galaxy clusters.

The gamma ray emission produced by $\chi \chi$ interaction 
extends up to energies corresponding to the neutralino mass (see Fig.11).
One way to disentangle a particular model for the cluster gamma-ray emission 
is provided by the future high-energy gamma-ray 
experiments. They will allow to observe galaxy clusters with 
a photon statistics sufficient
to disentangle the gamma-ray spectra produced by the  
$\pi^0 \rightarrow \gamma + \gamma $ electromagnetic 
decay, as predicted either in DM annihilation models ($\chi \chi \to ff,WW \to \pi^0$)
or in secondary electron models ($pp \to \pi^0$; see, e.g., 
Colafrancesco and Blasi 1998), from the gamma-ray emission produced by 
the bremsstrahlung of primary cosmic ray electrons.
These observational capabilities will be available, however, in the next future.
We will discuss more specifically this issue in a forthcoming paper.

Finally, we want to discuss a number of additional stringent predictions through which 
the model discussed here can be tested.
\newline
{\it (i)} According to the fact that DM is present in all large scale structures, 
we should observe radio halos in every cluster, which  is not actually observed.
In the framework of this paper,
we have shown that radio halos can be fitted with sensitively high 
magnetic fields (with central values $B_0 \sim 5 \div 100 ~\mu$G).
Then, only clusters which have such high magnetic fields can show  
a bright radio halo.
All the other clusters are expected to have faint radio halos 
that can brighten up when there is some effect that raises the magnetic field 
amplitude, for example, a merging event (see the numerical simulations of
K. Roettiger. 1999).
\newline
{\it (ii)} A possible problem could be given by the fact that cooling flow clusters, which have
usually high  magnetic field in their central cooling regions, do 
not show strong evidence for extended radio halos.
However, if one assumes that the magnetic field in cooling flow clusters is quite 
peaked near the cluster center and decreases rapidly towards the outskirts, 
the radio halo emission predicted in the present model 
is a factor $\sim 10^2\div 10^6$ lower than that of a 
non-cooling flow cluster with the same mass
[these estimates are obtained for a Coma-like cluster using 
the magnetic field profile given in eq.(19) with
parameters, $r_{c,B}=0.1 h_{50}^{-1}$ Mpc and $w=0.5\div 2$].
Nonetheless, it remains true that  our model does predict that cooling flow cluster
should possess small-size, low-luminosity and low-surface brightness radio halos.
\newline
{\it (iii)} Under the assumption that all clusters show a universal DM profile,
the radio halo spectrum is basically the same unless the magnetic 
field configuration of the cluster is very peculiar.
For example, steep radio halo spectra can be obtained mainly due to the presence of
magnetic fields that decrease strongly from the cluster center.
\newline 
Little is known about the presence and structure of the IC magnetic 
fields. Faraday rotation measurements (Vallee et al. 1986, 1987) give a lower estimate of the magnetic field in small
scale regions of the cluster (Dolag et al. 1999), 
and we could have to deal  with large scale IC magnetic fields 
that are lower than those ($B_0 \sim 5 \div 100$ $\mu$G, $B_{uniform} \sim 1 \div 3$ $\mu$G) 
required by the present model to explain radio halos.
However, even though the magnetic field are not so strong in clusters, 
the present model certainly yields a natural explanation for the origin of 
the seed 
high-energy electrons, which are a necessary input for any model of radio halo (and relic) formation.
The diffuse radio emission due to such a population of seed electrons could then be 
boosted by the amplification of the IC magnetic field subsequent to a strong cluster
merger (Roettiger 1999), or the seed electrons could be reaccelerated by intracluster turbulence
(Deiss et al. 1997, Eilek and Weatherall 1999, Brunetti et al. 1999).

While at the moment, we observe radio halos in more than 20 clusters 
at different redshifts, the definite test for the theory of the radio halo 
origin proposed here 
is committed to obtain a larger, unbiased survey of galaxy clusters through high 
radio sensitivity observations.
This search should be complemented with the search for gamma-ray emission from
galaxy clusters whose predicted intensity is matched by the sensitivities of 
the next generation gamma-ray experiments (GLAST, AGILE, MAGIC, VERITAS, ARGO, STACEE).

In conclusion,  we want to emphasize that the astrophysical expectations from 
the $\chi \chi$ annihilation are consistent, at the moment, 
with the constraints set by all the available 
observations on clusters containing radio halos.
The astrophysical and fundamental physics requirements on the model  
discussed here are stringent, but still well allowed. 
It is also appealing, in these respects,  to expect that some astrophysical features 
of galaxy clusters might give information on the fundamental properties of the DM 
particles.

\vskip 1.0truecm
\newline
{\bf Aknowledgments}. The authors acknowledge several stimulating discussions with 
P. Lipari, D. Fargion, A. Fabian and L. Feretti.

\clearpage
\begin{appendix}
\section{APPENDIX. The production spectrum of high-energy electrons}

In this Appendix we will derive an analytical approximation for the source spectrum 
of electrons
resulting from the decay products of $\chi \chi$ annihilation.
Here we will refer mainly to the analytical approaches of Rudaz and Stecker 
(1988; hereafter RS) 
and of Kamionkowski and Turner (1991; hereafter KT)
who gave analytical approximations of the positron source functions for models 
in which neutralinos
annihilate mainly into fermions ($\chi \chi \to f f $) 
or gauge bosons ($\chi \chi \to W W $) , respectively.
In the following, we derive the electron source spectra for generic values of the
neutralino number density, $n_{\chi}(r)$, and annihilation cross section, 
$\langle \sigma V \rangle_A$.

\subsection{Fermion dominated annihilation}
Following RS, we  consider three main sources of \el from $\chi \chi$ annihilation:
(P1) first generation, prompt electrons with a continuum spectrum;
(P2) second generation, prompt electrons;
($\pi$) electrons produced from the decay of $\pi ^{\pm}$.
The electron source spectrum, $Q(E,r)$, can be written as the sum of 
these three components
\be
Q_{ff}(E,r) = Q_{P1}(E,r) + Q_{P2}(E,r) + Q_{\pi}(E,r) ~.
\ee
For a neutralino with mass $M_{\chi}=100$ GeV,
the first generation source spectrum takes contributions  
mainly from $b  \to e, c  \to e, \tau^{\pm} \to e^{\pm}$ (see RS).
Taking into account the different energy distributions of the previous decays, the first generation
electron spectrum is found to be well approximated by
\be
Q_{P1}(E,r) = n^2_{\chi} \langle \sigma V\rangle_A \zeta_1 f_1(E)
\ee
with $\zeta_1 \approx 0.6/(kM_{\chi})$ and $k \approx 0.7$.
\newline
The function $f_1(E)$ is given by 
\be
f_1(x) = {0.17 \over 31} f_{\tau}(x) + 0.13 \times {3 \over 31} f_c(x) + 0.1 \times {27 \over 31} 
f_b(x)
\ee
where
\be
f_{\tau}(x) = f_b(x) = {5 \over 3} + {4 \over 3} x^3 - 3 x^2
\ee
and
\be
f_c(x) = 2\bigg( 1 + 2 x^3 - 3 x^2 \bigg)
\ee
in terms of the adimensional quantity $x = E/M_{\chi}$.
In eq.(A3) we replace the values assumed by RS for the leptonic decay branching ratios with the
present values taken from the Particle Data Group (Groom et al. 2000).
Neglecting the differences between $f_{\tau}, f_b$ and $f_c$, the source function $Q_{P1}$ -- as noticed by
RS -- can be approximated by 
\be
Q_{P1}(E,r) \approx 0.11 n^2_{\chi} \langle \sigma V\rangle_A \cdot  f(E)
\ee
where
\be
f(E) = {1 \over k M_{\chi}} \theta( k M_{\chi} - E)
\ee
where the function $\theta( k M_{\chi} - E)$ is the Heaviside function.
The source function $Q_{P1}(E,r)$
is basically constant up to $E \approx k M_{\chi}$ and then drops to zero for
higher energies (see Fig.1).

\newline
The second generation source spectrum 
(arising from the leptonic decays in the final states of the first generation
decays)
is found to be well approximated by:
\be
Q_{P2}(E,r) = n^2_{\chi} \langle \sigma V\rangle_A \zeta_2 f_2(E)
\ee
with
\be
f_2(E) = {1 \over k^2 M_{\chi}} ln \bigg[{ k^2 M_{\chi} \over E} \bigg]
\ee
where, again, $k \approx 0.7$ and 
\be
\zeta_2 \approx {0.17 \over 31} + 0.13 \times  {3 \over 31} + 0.1 \times 2 \times 
{27 \over 31} ~.
\ee
This source function goes rapidly to zero for 
$E \rightarrow k^2 M_{\chi} \approx M_{\chi} /2$ (see Fig.1).
Again, in eq.(A10)  we replace the values assumed by RS for the leptonic decay branching ratios with the
present values taken from the Particle Data Group (Groom et al. 2000).

\newline
Following RS, we find that the source spectrum from $\pi^{\pm}$ decay is found to be well approximated by:
\be
Q_{\pi}(E,r) = n^2_{\chi} \langle \sigma V\rangle_A \zeta_{\pi} f_{\pi}(E)
\ee
with 
\be
f_{\pi}(E) = e^{-0.68 E} + 0.115 e^{-0.276 E}
\ee
and 
$\zeta_{\pi} \approx 1.74 (100 ~{\rm GeV}/ M_{\chi})$.


\noindent
The three source functions for P1, P2 and $\pi$ contributions together with the total 
source function are plotted in Figs.1 and 2.

\subsection{Gauge boson dominated annihilation}

In addition to the electron produced by the direct decays of vector  bosons 
(typical energy of $\sim M_{\chi}
/ 2$) there is a continuum spectrum of electrons at 
$E\simlt M_{\chi}/2$ which are produced as
secondary decay products ($W^{\pm} \to \tau^{\pm} \to e^{\pm}$,
$W^{\pm} \to c \to e^{\pm}$) and from the decays of charged pions produced in hadronic decays
of the $W^{\pm}$ and $Z^0$ bosons.
Since the electron distributions arising from the $W$ and the $Z$ bosons
are qualitatively similar, we will restrict our analysis to the $W$ case.
Following KT, the source distribution function of secondary-decay electrons, integrated over the
quark and lepton energies, is
\be
Q_{W,\mu,\tau,c}(E,r)= n^2_{\chi} \langle \sigma V\rangle_A \zeta_{W,\mu,\tau,c} f_{W,\mu,\tau,c}(E)
\ee
where
\be
\zeta_{W,\mu,\tau,c} = B_{W\to \mu} + 0.17 B_{W\to \tau} + 0.13 B_{W\to c}
\ee
and
\be
f_{W,\mu,\tau,c}(E) = 
\left\{
        \begin{array}{l l}
		{1 \over kM_{\chi} \delta} ln\bigg[{(1+\delta) \over (1- \delta)}\bigg] 
		~~~~~ &
              \mbox{for $E \leq {kM_{\chi} \over 2} (1+\delta)$},\\
		{1 \over kM_{\chi} \delta} ln\bigg[{k  M_{\chi}(1+\delta) \over 2E}\bigg] 
		~~~~~ &
              \mbox{for ${kM_{\chi} \over 2} (1+\delta) \leq E \leq {kM_{\chi} \over 2} (1+\delta)$},\\
		0		~~~~~ &
              \mbox{for $E \geq {kM_{\chi} \over 2} (1+\delta)$},
        \end{array}
        \right\}
\ee
In this expression, $\delta = [1 - (M_W/ M_{\chi})^2]^{1/2}$ with $M_W = 80$ GeV being the W boson mass.
Adopting values $B_{W\to \mu} = 0.11$, $B_{W\to \tau} = 0.11$ and $B_{W\to c} = 0.34$
for the previous branching ratios, we obtain
$\zeta_{W,\mu,\tau,c} \approx 0.1729$. 

The hadronization of quarks from gauge boson decays results in a shower of charged pions which
eventually decay into electrons ($\pi^{\pm} \to \mu^{\pm} \to e^{\pm}$).
Following KT, the source function for the pion-produced electrons, integrated over the quark energy
distribution, writes
\be
Q_{W,\pi}(E,r)= n^2_{\chi} \langle \sigma V\rangle_A \zeta_{W,\pi} f_{W,\pi}(E)
\ee
where
$\zeta_{W,\pi} \approx 2/3$ and
\be
f_{W,\pi}(E) = {1 \over M_{\chi} \delta} \int_{E_{min}}^{E_{max}} dE' 
\bigg[ 93 e^{-68 E/E'} + 56 e^{-27.6 E/E'} \bigg]
\ee
where $E_{min}=M_{\chi}(1 - \delta)/2$ and $E_{max}=M_{\chi}(1 + \delta)/2$.
The total source function for this case writes
\be
Q_{WW}(E,r)=Q_{W,\mu,\tau,c}(E,r) + Q_{W,\pi}(E,r)
\ee
and is compared to the source function $Q_{ff}(E,r)$ in Figs.1 and 2.

\end{appendix}

%

\clearpage
\begin{figure}[h]
 \begin{center}
\mbox{\epsfig{file=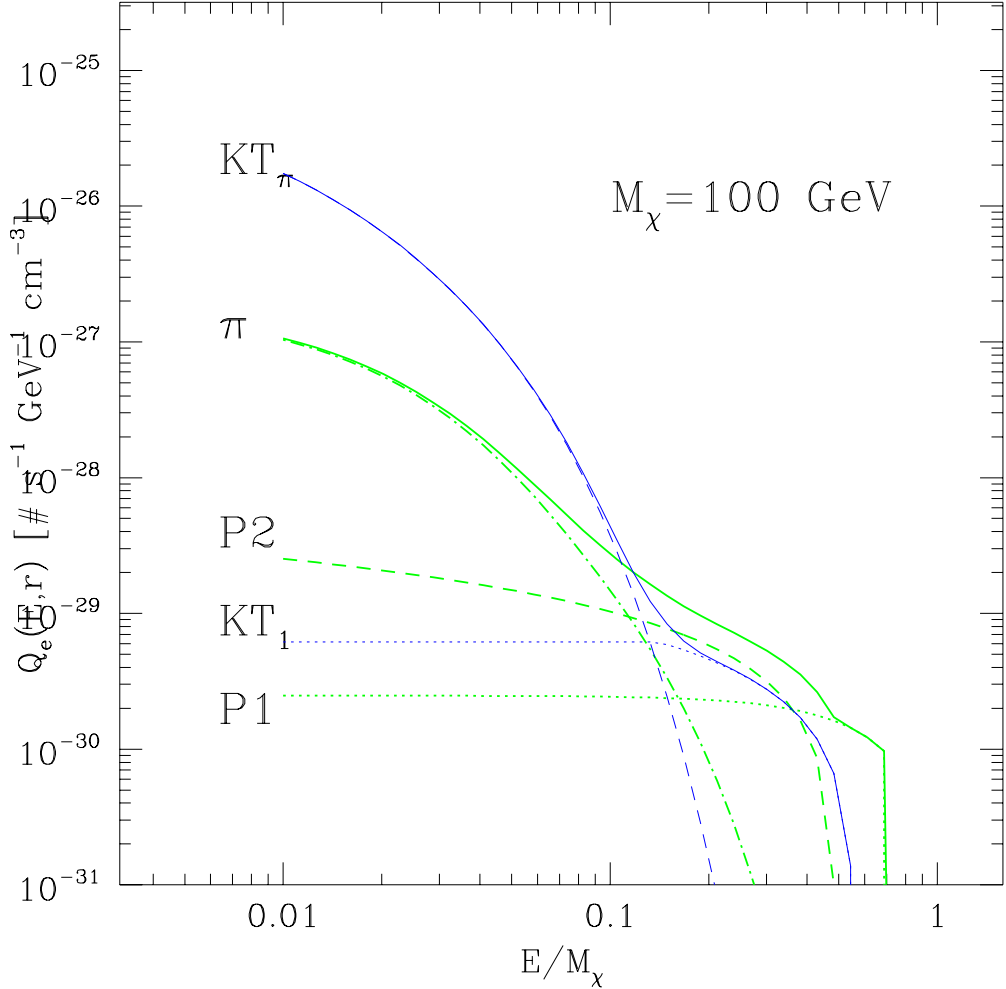,height=14.cm,width=14.cm,angle=0.0}}
 \end{center}
\caption{The source spectrum for a model in which fermions dominate the 
neutralino annihilation products ($\chi \chi \to ff$): 
first generation, prompt electrons (P1),
second generation, prompt electrons (P2) and secondary electrons produced in the decay
of charged pions ($\pi$).
The light solid curve is the total source spectrum for this case.
\newline
We also show the source spectra for the model in which gauge bosons dominate the 
annihilation. 
The contributions from the decays ($W^{\pm} \to \tau^{\pm} \to e^{\pm}$, 
$W^{\pm} \to c \to e^{\pm}$) and from the decays of charges pions produced in hadronic
decays of $W^{\pm}$ and $Z^0$ bosons are labelled KT$_1$ (light dotted curve) and  
KT$_{\pi}$ (light dashed curve), respectively.
The light solid curve is the total source spectrum for the case
($\chi \chi \to WW$). 
A neutralino density $n_{\chi} = 1$ cm$^{-3}$ and annihilation cross section 
$\langle \sigma V \rangle_A= 10^{-27}$ cm$^3$ s$^{-1}$ have been used in this plot.
}
\label{fig1}
\end{figure}

\clearpage
\begin{figure}[h]
 \begin{center}
\mbox{\epsfig{file=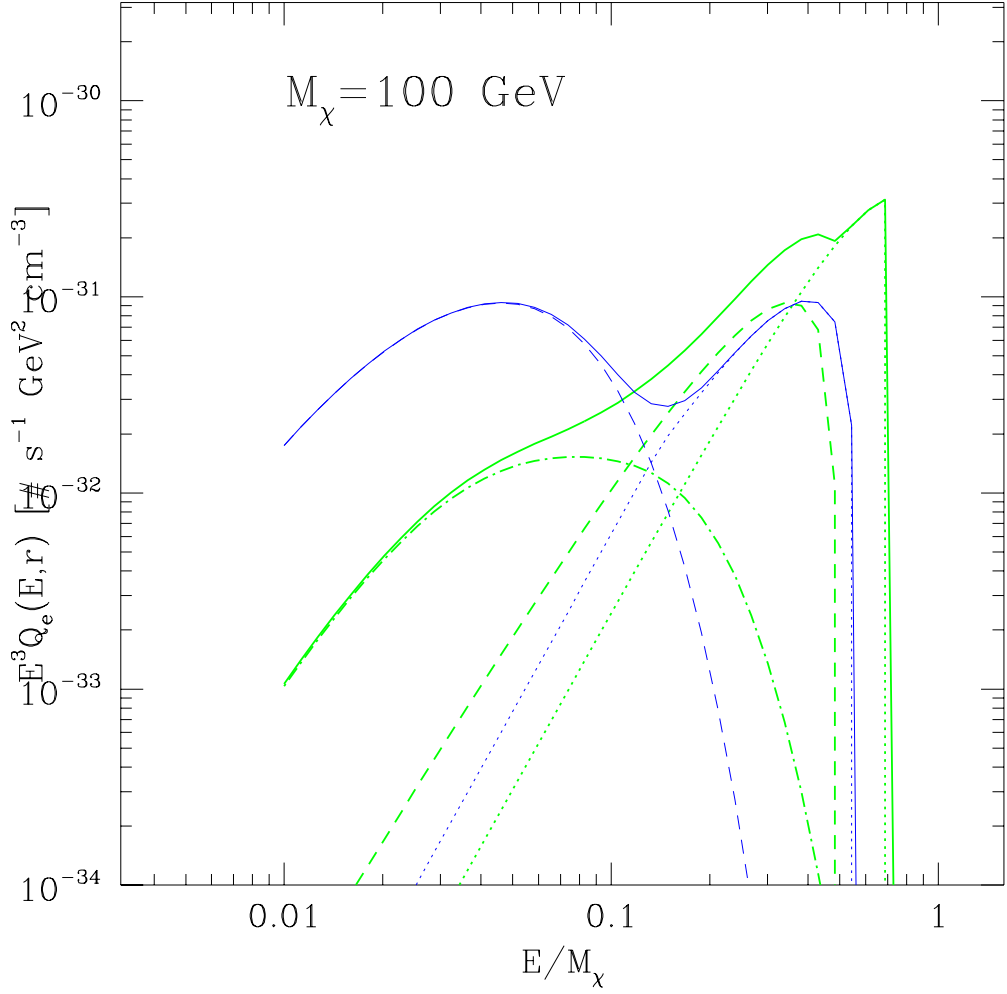,height=14.cm,width=14.cm,angle=0.0}}
 \end{center}
\caption{The quantity $E^3 Q_e(E,r)$ is shown for the source spectra plotted in Fig.1.
Curves are labelled as in Fig.1.
}
\label{fig2}
\end{figure}

\clearpage
\begin{figure}[h]
 \begin{center}
\mbox{\epsfig{file=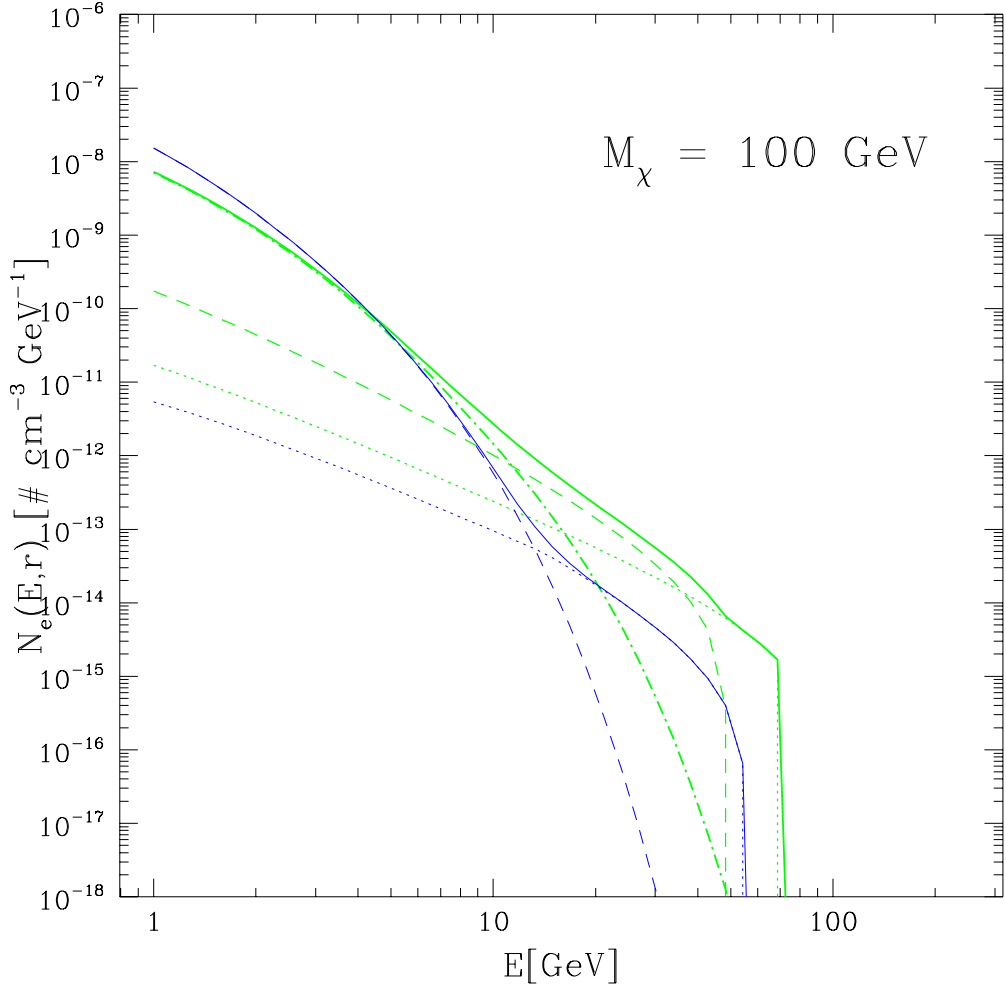,height=14.cm,width=14.cm,angle=0.0}}
 \end{center}
\caption{The overall equilibrium spectrum $N_e(E,r)$ of the electrons 
whose source functions are shown in Fig. 1.
The equilibrium spectra are calculated separately for each source electron population
given in Fig.1.
The heavy and light solid curves are the total equilibrium spectra for the cases 
$\chi \chi \to f f $ and $\chi \chi \to WW $, respectively.
A neutralino with mass $M_{\chi}=100$ GeV has been adopted.
}
\label{fig3}
\end{figure}

\clearpage
\begin{figure}[h]
 \begin{center}
\mbox{\epsfig{file=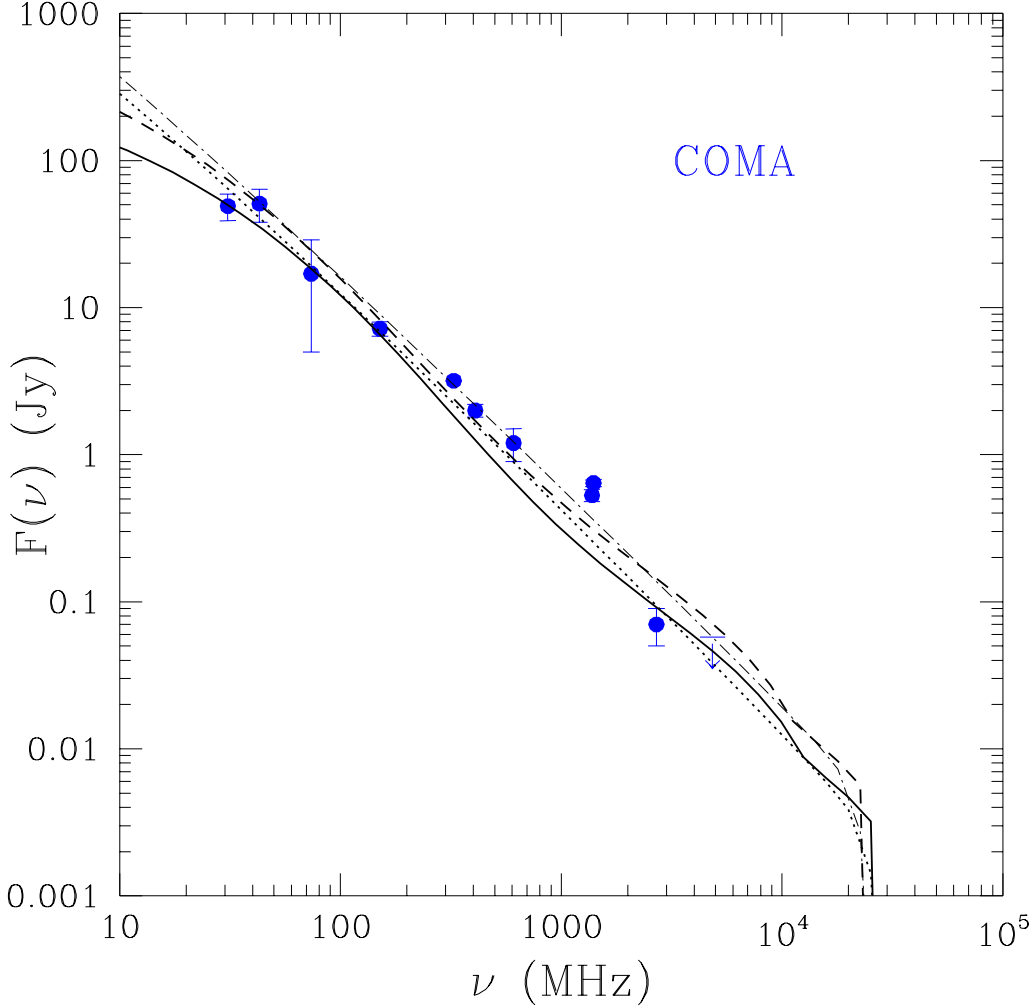,height=14.cm,width=14.cm,angle=0.0}}
 \end{center}
\caption{The Coma radio halo spectrum predicted in a model in which $\chi$
annihilation is dominated by fermions ($\chi \chi \to ff$).
The curves are for a uniform magnetic field of $B_{uniform} = 1.3 \mu$G (dashed) 
and for a radially
decreasing magnetic field with central value $B_0 = 8 \mu$G (solid).
We also show the radio halo spectra obtained from a power-law approximation, 
$Q_e \sim E^{-1.9}$ to the true source spectrum shown in Fig.1 with $B_{uniform}=1.3
\mu$G (dot-dashed) and $B_0=8 \mu$G (dotted).
A constant core profile with $r_c=0.4 h_{50}^{-1}$ Mpc, $\beta=0.76$ has been adopted.
The radio halo emission has been integrated out to $1.3h_{50}^{-1}$ Mpc.
Data are taken from Deiss et al. (1997).
}
\label{fig4}
\end{figure}

\clearpage
\begin{figure}[h]
 \begin{center}
\mbox{\epsfig{file=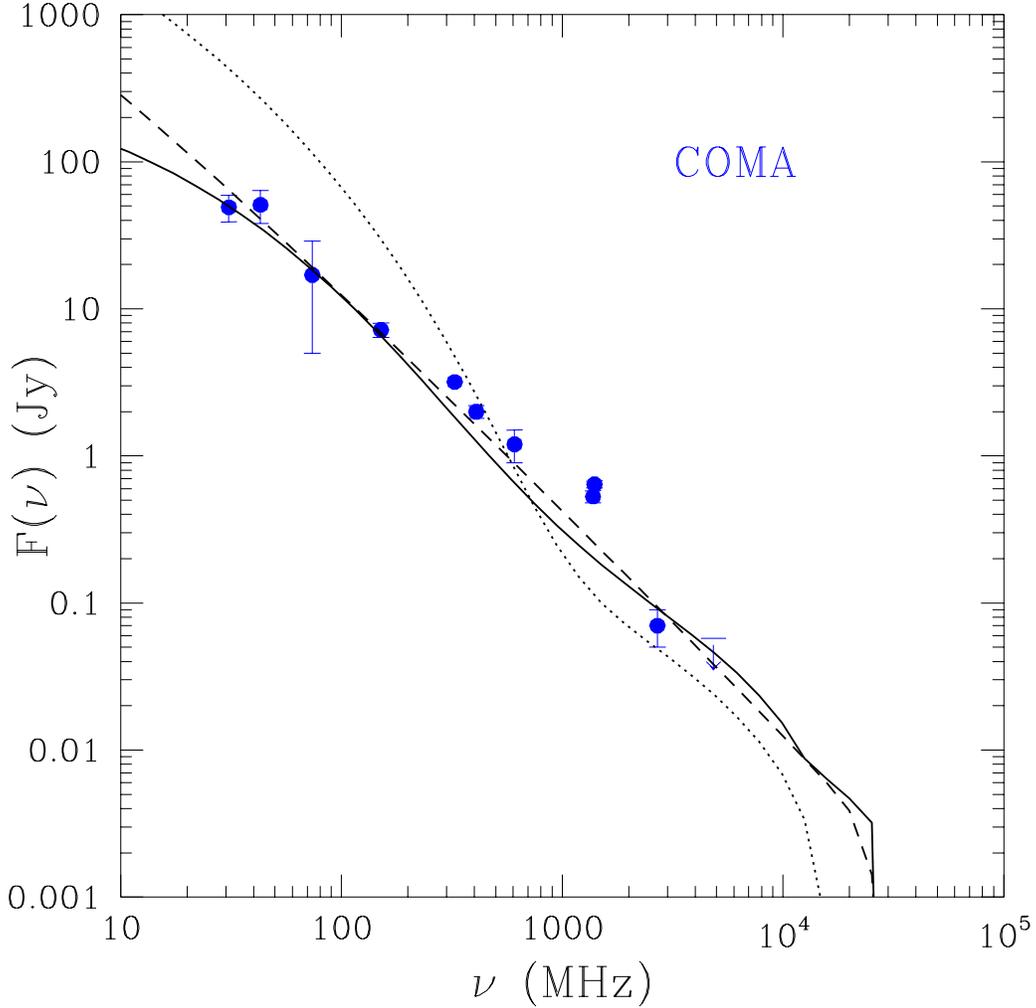,height=14.cm,width=14.cm,angle=0.0}}
 \end{center}
\caption{The Coma radio halo spectrum predicted in a model in which $\chi$
annihilation is dominated by fermions ($\chi \chi \to ff$;  solid curve) 
is compared to the model in which annihilation is dominated 
by higgsinos ($\chi \chi \to WW$;  dotted curve).
The power-law approximation for the case $\chi \chi \to ff$ is also shown (dashed curve).
The curves are plotted for a radially
decreasing magnetic field with central value $B_0 = 8 \mu$G.
A constant core profile with $r_c=0.4h_{50}^{-1}$ Mpc, $\beta=0.76$ has been adopted.
The radio halo emission has been integrated out to $1.3h_{50}^{-1}$ Mpc.
Data are taken from Deiss et al. (1997).
}
\label{fig5}
\end{figure}

\clearpage
\begin{figure}[h]
 \begin{center}
\mbox{\epsfig{file=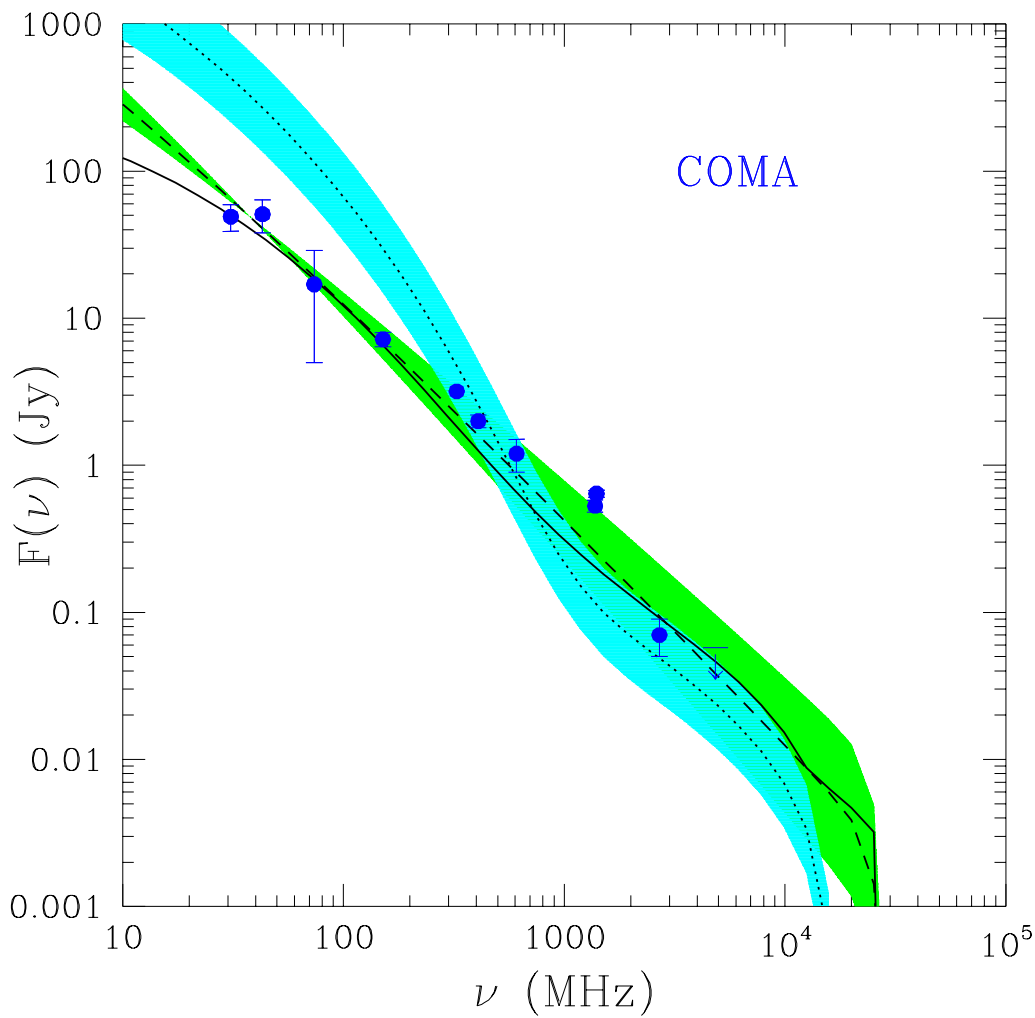,height=14.cm,width=14.cm,angle=0.0}}
 \end{center}
\caption{In this figure we show how the uncertainties in the electron source spectrum 
reflect on the shape of the radio halo spectrum of Coma.
Shaded areas enclose radio halo spectra evaluated considering both an uncertainty of 
a factor $\pm 2$ in normalization and $\pm 10 \%$ in the slope of the source spectra.
Dark gray region refers to the
$\chi \chi \to ff$  case and pale gray region to the
$\chi \chi \to WW$ case.
Curves are labelled as in Fig.5.
}
\label{fig6}
\end{figure}

\clearpage
\begin{figure}[h]
 \begin{center}
\mbox{\epsfig{file=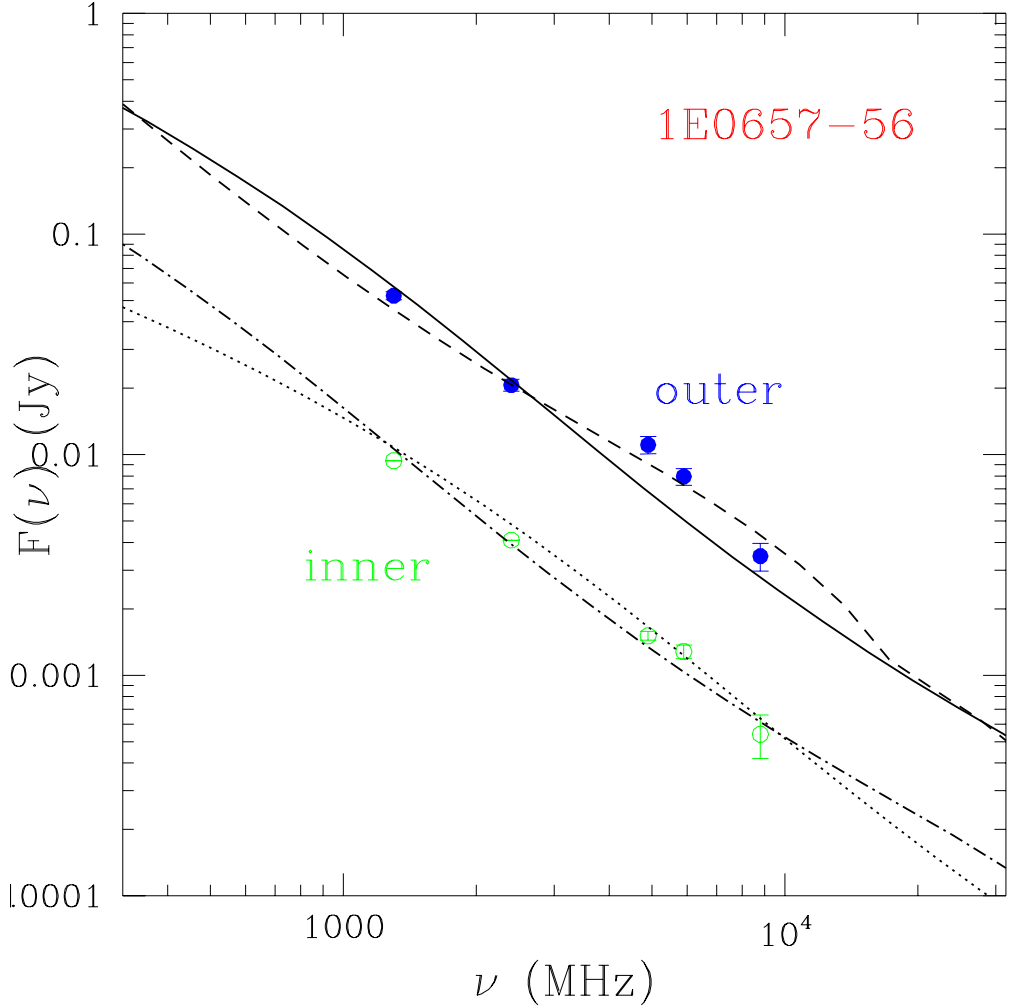,height=14.cm,width=14.cm,angle=0.0}}
 \end{center}
\caption{The radio halo spectrum of the cluster 1E0657-56 
in the inner (empty circles) and outer (filled circles) regions.
Dashed and dot-dashed curves are for a uniform field $B_{uniform} =2 \mu$G  and 
$B_{uniform} = 9 \mu$G, respectively.
Solid and dotted curves are for a magnetic field with a radial dependence
$B=B_0\bigg[ 1 + (r/r_{c,B})^2 \bigg]^{-0.5}$
with a central amplitude $B_0 = 100 \mu$G and $B_0 = 90 \mu$G, respectively.
Curves are shown for a model in which neutralino annihilation is dominated by fermions
($\chi \chi \to f f $).
Constant core density profiles have been adopted here with parameters
$R_{halo}=2 h_{50}^{-1}$ Mpc, $r_{c}=0.38 h_{50}^{-1}$ Mpc, $\beta=0.7$ and $n_{\chi,0}
=9 \cdot 10^{-3}$ cm$^{-3}$ for the outer region and
$R_{halo}=0.6 h_{50}^{-1}$ Mpc, $r_{c}=0.08 h_{50}^{-1}$ Mpc, $\beta=0.49$ and $n_{\chi,0}
=9 \cdot 10^{-3}$ cm$^{-3}$ for the inner region (see Liang et al. 2000).
Data are from Liang et al. (2000).
}
\label{fig7}
\end{figure}

\clearpage
\begin{figure}[h]
 \begin{center}
\hbox{
\epsfig{file=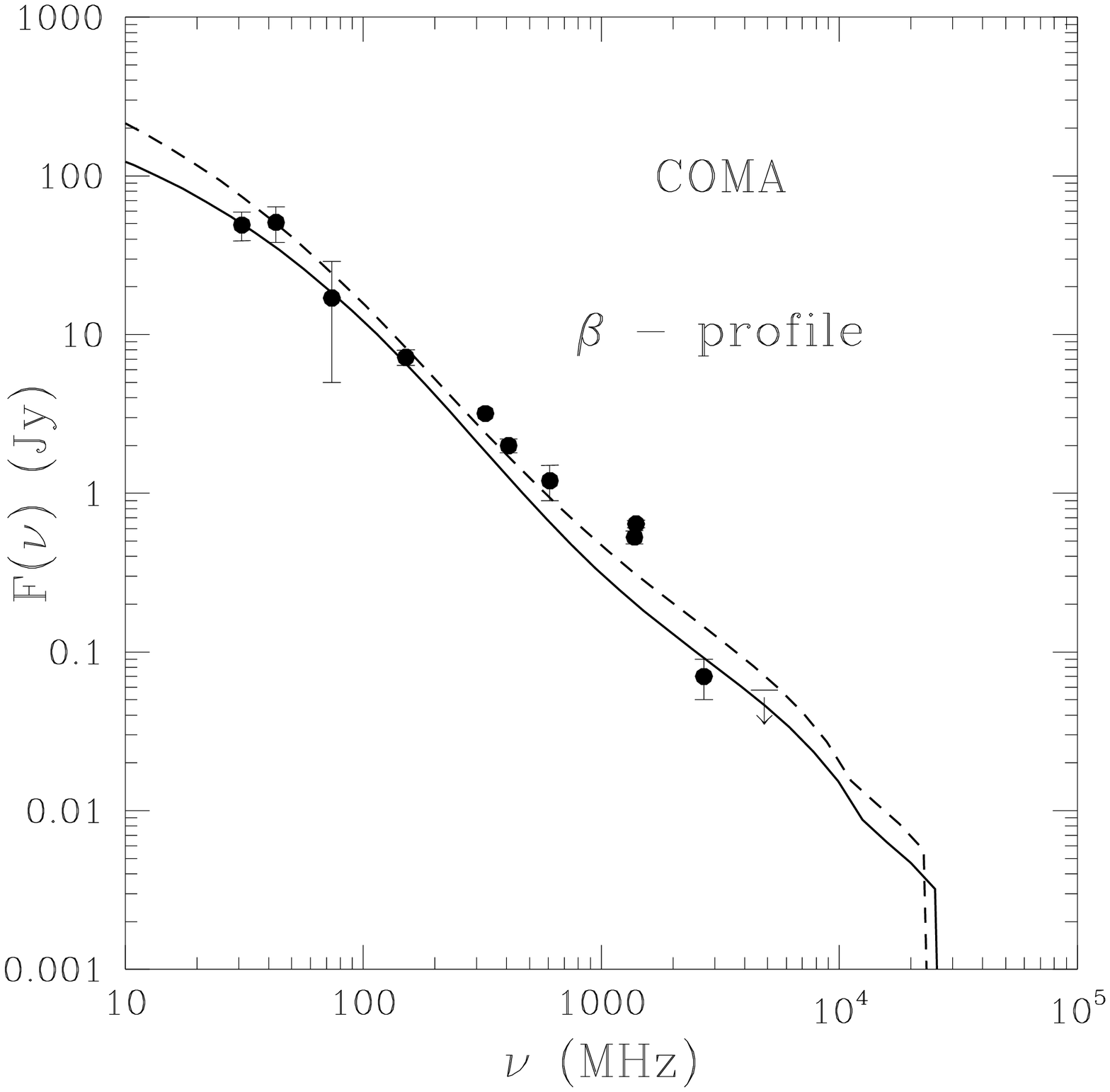,height=8.cm,width=8.cm,angle=0.0}
\epsfig{file=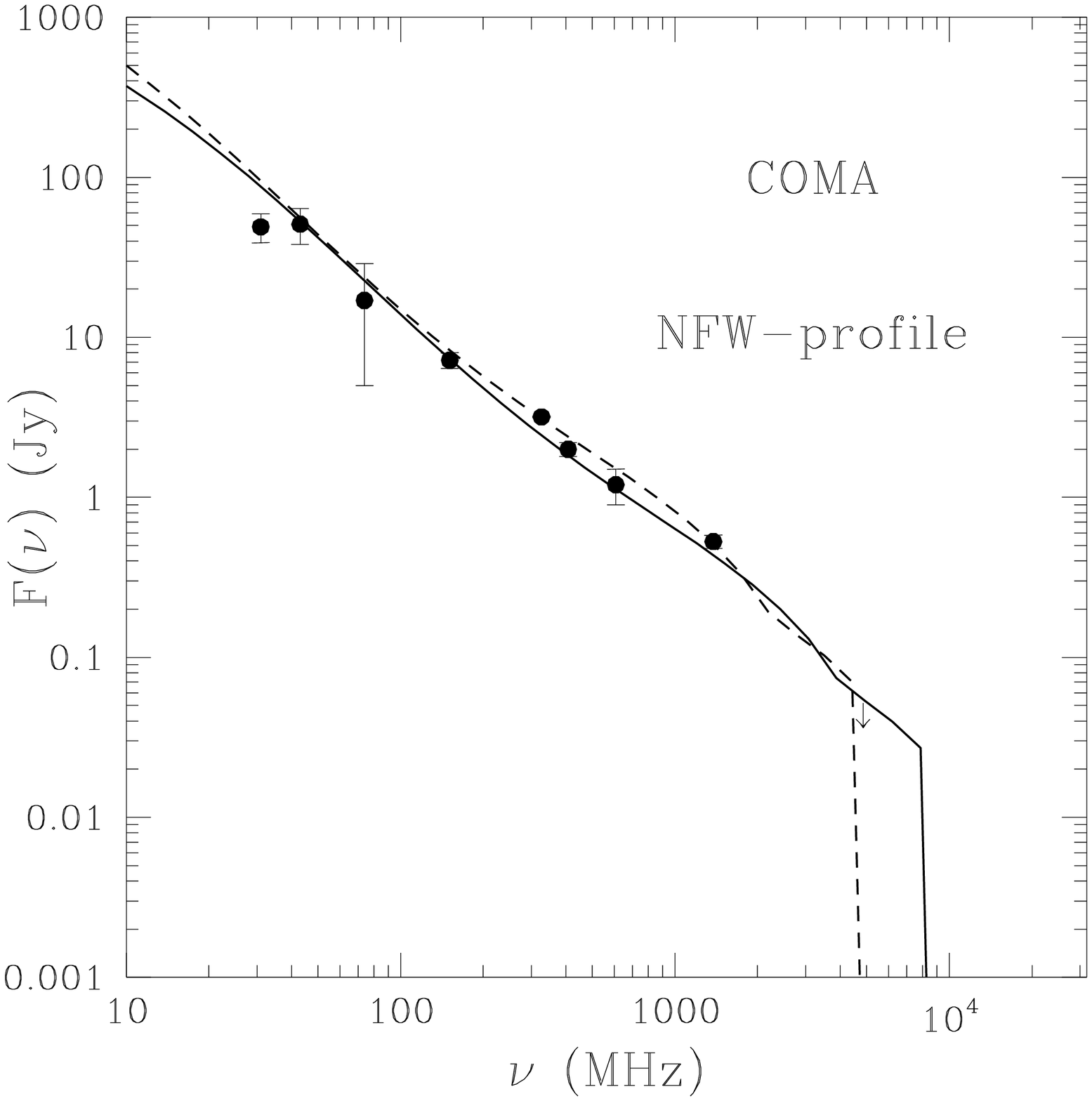,height=8.cm,width=8.cm,angle=0.0}
}
 \end{center}
\caption{The Coma radio halo spectrum for a  constant core beta-profile (left panel) 
and for a NFW profile (right panel).
Dashed and solid curves refer to the $B_{uniform}=1.3 \mu$ G and $B_0=8 \mu$G
for the beta-profile (left panel) and to $B_{uniform}=0.45 \mu$ G and $B_0=1.4 \mu$G
for the NFW profile (right panel).
A neutralino annihilation model $\chi \chi \to ff$ has been assumed.
Data are from Deiss et al. (1997).
}
\label{fig8}
\end{figure}

\clearpage
\begin{figure}[h]
 \begin{center}
\mbox{\epsfig{file=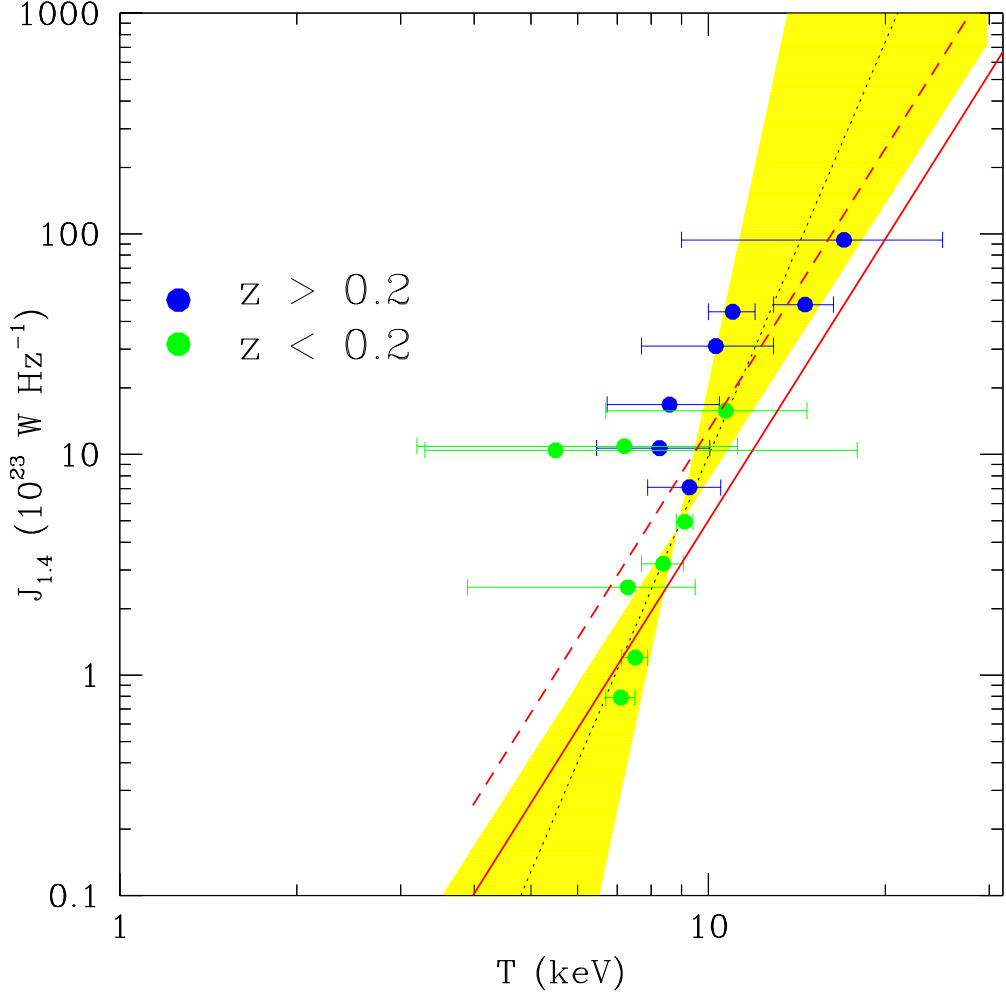,height=14.cm,width=14.cm,angle=0.0}}
 \end{center}
\caption{The $J_{1.4}-T$ correlation is shown for the radio halo clusters 
at $z > 0.2$ (dark dots)  and at $z<0.2$ (gray dots).
The dotted line is the best fit power-law to the data 
(the shaded area contain the uncertainty region of the best fit points, see Section
6 for details).
The heavy lines are the relations $J_{1.4} \sim T^{4.25}$ expected in the neutralino
annihilation model discussed in the paper and are 
evaluated at $z=0$ (solid) and at $z=0.25$ (dashed), respectively.
data are taken from Feretti (1999), Giovannini et al. (1999), Liang (1999), Owen
et al. (1999).
}
\label{fig9}
\end{figure}

\clearpage
\begin{figure}[h]
 \begin{center}
\mbox{\epsfig{file=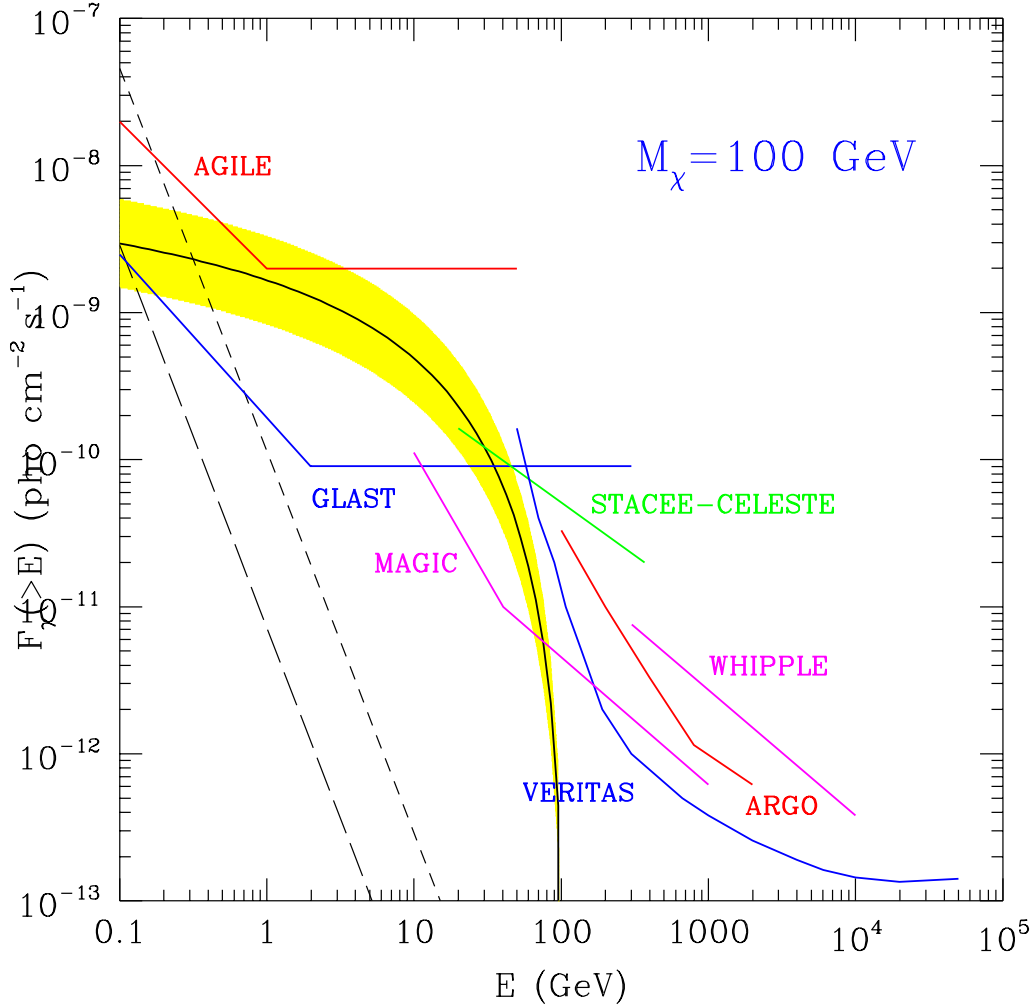,height=14.cm,width=14.cm,angle=0.0}}
 \end{center}
\caption{The gamma-ray emission predicted from $\chi \chi$ annihilation in Coma 
(solid curve) is compared to the sensitivities of the operating and planned
gamma ray experiments.
A neutralino with mass $M_{\chi}=100$ GeV has been assumed.
The shaded area shows the uncertainty in the gammma-ray spectrum due to an uncertainty
of a factor $2$ in the normalization of the $\pi^0$ source spectrum (see Section 3
for details).
The short and long dashed curves refer to the gamma ray emission produced by
relativistic bremmstrahlung of a population of primary cosmic rays and are calculated
using the formula given in Sreekumar et al. (1996) 
for two choices of the Coma magnetic field $B= 0.3 \mu$G and $B= 1 \mu$G, respectively. 
}
\label{fig10}
\end{figure}

\clearpage
\begin{figure}[h]
 \begin{center}
\mbox{\epsfig{file=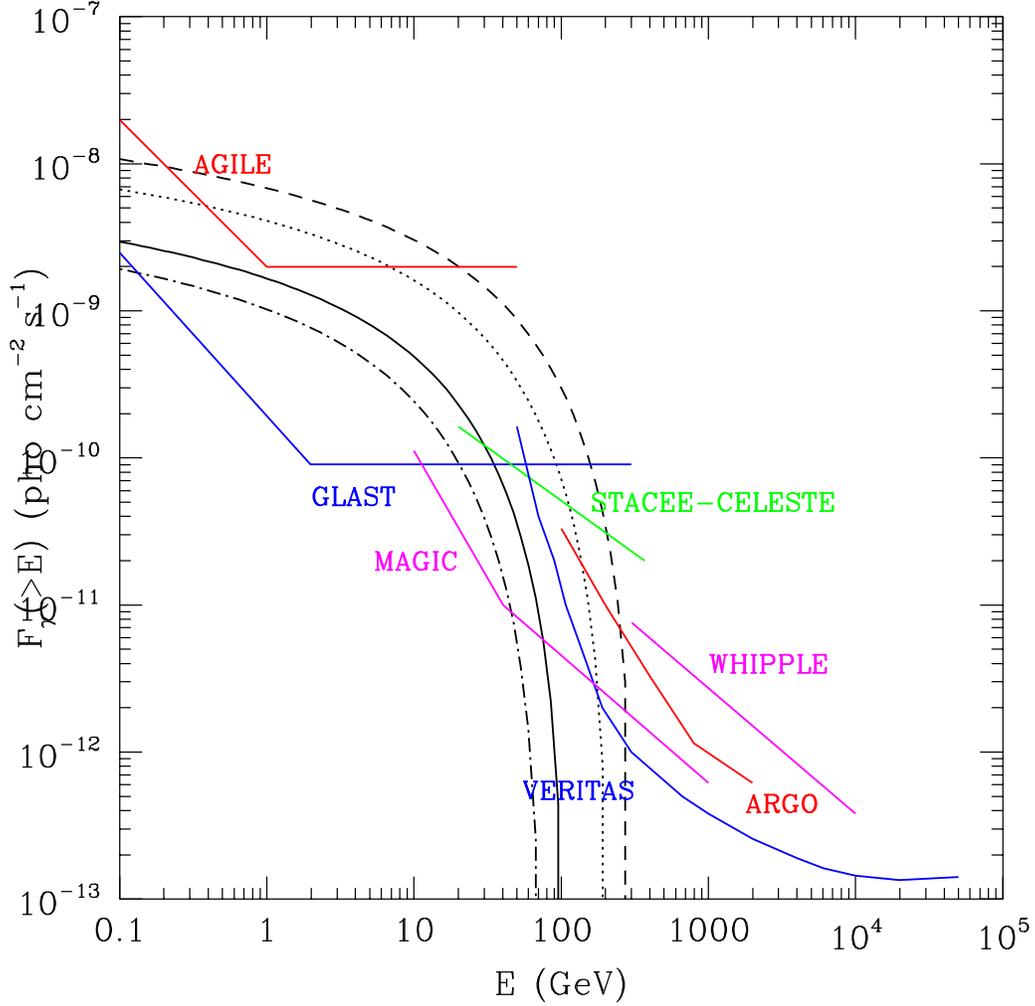,height=14.cm,width=14.cm,angle=0.0}}
 \end{center}
\caption{The gamma-ray emission predicted from $\chi \chi$ annihilation in galaxy
clusters (solid curve) calculated for different neutralino masses:
$M_{\chi}=70$ GeV (dot-dashed curve), 
$M_{\chi}=100$ GeV (solid curve),
$M_{\chi}=200$ GeV (dotted curve)
and  $M_{\chi}=300$ GeV (dashed curve).
The sensitivities of the operating and planned gamma ray experiments are also shown.
Combined gamma ray observations of clusters from $\sim 1$ GeV to $500$ GeV can clearly 
determine the neutralino mass from both the intensity of the spectrum and its
high-energy cutoff (see Section 7 for details).
}
\label{fig11}
\end{figure}

\end{document}